\documentclass[letterpaper,twocolumn,10pt]{article}
\usepackage{usenix-2020-09}

\usepackage{tikz}
\usepackage{amsmath}

\usepackage{filecontents}

\usepackage{booktabs} % For formal tables
\usepackage{mathrsfs}

\usepackage{hyperref}
\usepackage{multirow}
\usepackage[utf8]{inputenc} 
\usepackage[ruled, lined, longend, linesnumbered]{algorithm2e}

\usepackage{array}
\usepackage[T1]{fontenc}
\usepackage{url}
\usepackage{gensymb}

\usepackage{tikz}
\newcommand\encircle[1]{%
  \tikz[baseline=(X.base)] 
    \node (X) [draw, shape=circle, inner sep=0, fill=black, text=white] {\strut #1};%
}

\usepackage{tikz}
\newcommand\Encircle[1]{%
  \tikz[baseline=(X.base)] 
    \node (X) [draw, shape=circle, inner sep=0, fill=white, text=black] {\strut #1};%
}

\usepackage{wrapfig}
\usepackage{comment}
\usepackage{colortbl}
\usepackage{float}
\usepackage{tabularx}

\usepackage{color,soul}
\usepackage{graphics}
\usepackage{graphicx}
\usepackage{lipsum}

\usepackage{CJKutf8} % JAP, CHN, KOR

\usepackage[inline]{enumitem}	% for fine control of list env. support inline enumerate. 
\usepackage{listings} % for code listing
\usepackage{lipsum}

\usepackage{capt-of}	% xzl: used to override the ``figure'' and ``table'' in caption.

\usepackage{ulem}  % xzl: for strikeout, which seems more robust than \st{} provided by soul package. 

\ifdefined\HCode
\usepackage[compatibility=false]{caption}

\else
\usepackage{subcaption} 
\fi

\usepackage{mathrsfs}	% xzl: for Ralph Smith’s Formal Script Font (rsfs)
\usepackage{amsfonts}

\usepackage[export]{adjustbox} % xzl: flush figure to left/right

\definecolor{myForestGreen}{RGB}{34, 139, 34}

\definecolor{refkey}{RGB}{34,139,34}
\definecolor{labelkey}{rgb}{0,0,1}

\renewcommand{\paragraph}[1]{\vskip 3pt\noindent\textbf{#1 }}	 % used to be 6pt

  {\begin{list}{$\bullet$}%
     {\setlength{\parsep}{0pt}%
      \setlength{\topsep}{0pt}%
      \setlength{\itemsep}{2pt}}}%
  {\end{list}}
\newcommand\Note[1]{\sethlcolor{green} \hl{#1}} % highlighted notes of other colors.

\newcommand\notettx[1]{\sethlcolor{orange} \hl{#1}} % highlighted notes of other colors.

\newcommand\sect[1]{Section~\ref{sec:#1}}	% NB: does not play well with \note{}
		
\newenvironment{myitemize}%
  {\begin{itemize}
	[leftmargin=0cm,
		itemindent=.3cm,
		labelwidth=\itemindent,
		labelsep=0pt,
		parsep=3pt,
		topsep=2pt,
		itemsep=1pt,
		align=left]
  }%
  {\end{itemize}}    

  {\begin{enumerate}
	[leftmargin=0cm,itemindent=.5cm,labelwidth=\itemindent,
		labelsep=0pt,
		parsep=1pt,
		topsep=1pt,
		itemsep=3pt,
		align=left]
  }%
  {\end{enumerate}}    
  
  {\begin{enumerate*}
	[label=\roman*)]
  }%
  {\end{enumerate*}}      

  {\begin{enumerate*}
	[label=\roman*)]
  }%
  {\end{enumerate*}}

\newcommand{\sys}{\textsc{Clique}}

\newcommand{\stp} {$\mathbb{P}$ }

\newcommand{\primary}{starter}
\newcommand{\prim}{starter}

\newcommand{\cube}{spatiotemporal cell}
\newcommand{\cubes}{spatiotemporal cells}

\makeatletter
\def\@copyrightspace{\relax}
\makeatother

\begin{document}
    \pagestyle{plain}
    
    % \title{Reinventing Query Processing Engines for Object Re-identification}
%    \title{\sys{}: A Video Engine for Object Re-identification at the City Scale}
    \title{\sys{}: Spatiotemporal Object Re-identification at the City Scale}

%	\author{Anonymous Authors}

	\author{
		{\rm Tiantu Xu}\\
		Purdue ECE \\
		\and
		{\rm Kaiwen Shen}\\
		Purdue ECE
		\and
		{\rm Yang Fu}\\
		UIUC
		\and
		{\rm Humphrey Shi}\\
		University of Oregon
		\and
		{\rm Felix Xiaozhu Lin}\\
		University of Virginia
	} % end author
	
%	\author[1]{Tiantu Xu}
%	\author[1]{Kaiwen Shen}
%	\author[2]{Yang Fu}
%	\author[3]{Humphrey Shi}
%	\author[4]{Felix Xiaozhu Lin}
%	\affil[1]{Purdue ECE} 
%	\affil[2]{University of Illinois at Urbana-Champaign}
%	\affil[3]{University of Oregon}
%	\affil[4]{University of Virginia}

    \maketitle
    
    % Abstract
    % !TeX root = main.tex
\begin{abstract}
Object re-identification (ReID) is a key application of city-scale cameras. 
While classic ReID tasks are often considered as image retrieval, 
we treat them as spatiotemporal queries for locations and times in which the target object appeared. 
Spatiotemporal reID is challenged by the accuracy limitation in computer vision algorithms and the colossal videos from city cameras. 
We present \sys{}, a practical ReID engine that builds upon two new techniques: 
(1) \sys{} assesses target occurrences by clustering fuzzy object features extracted by ReID algorithms, with each cluster representing the general impression of a distinct object to be matched against the input; 
(2) to search in videos, \sys{} samples cameras to maximize the spatiotemporal coverage and incrementally adds cameras for processing on demand.
Through evaluation on 25 hours of videos from 25 cameras, \sys{} reached a high accuracy of 0.87 (\textit{recall at 5}) across 70 queries and runs at 830$\times$ of video realtime in achieving high accuracy. 
\end{abstract}

\section{Introduction}
\label{sec:intro}

%Video analytics has see wide use in smart cities. 
%Catering to the need of computer vision (CV) aided tasks designed as 6future smart city solutions and the falling cost of camera hardware, surveillance cameras have proliferated in major big cities since the past decade.
%By the end of 2021, the total number of surveillance cameras in the world is expected to surpass 1 billion~\cite{billioncamreas, billioncamreas-2}. 

\paragraph{City-scale camera deployment}
As video intelligence advances and camera cost drops, 
city cameras expand fast. 
Strategically deployed near key locations, such as highway entrances or road intersections, 
multiple cameras (reported to be 2--5 per location~\cite{cityflow,visionzero}) offer complementary, often overlapped viewpoints of scenes. 
% For instance, Microsoft's camera deployment in Bellevue, WA~\cite{visionzero} places at least two cameras per traffic intersection; 
% This is also confirmed by popular city video datasets~\cite{cityflow}.

\paragraph{Object ReID on city videos}
A key application of city cameras is object re-identification (ReID): 
given an input image of an object X, searching for occurrences of X in a video repository. 
ReID has been an important computer vision task, seeing popular use cases such as crime investigation and  traffic planning~\cite{reidcases, visionzero,aicity}. 
Many ReID algorithms are proposed recently, fueled by neural networks~\cite{reid-survey, sun18eccv,zheng15iccv,zhong17cvpr,fu19aaai,tan2019cvpr,huang2019cvpr,fu19iccv}. 
Object ReID over city videos is typically ``finding a needle in haystack''.
%users query a repository of long videos produced by many cameras; 
The queried videos are long and produced by many cameras; 
the videos may not contain the input image, or any images from the camera that produced the input image (called the \textit{origin} camera);
the occurrences of target object can be rare and transient.
For instance, in a popular dataset of city traffic videos~\cite{cityflow}, 
% half of the vehicles only appear for less than 5 seconds. 
99\% of vehicles only appear for less than 10 seconds.
% 47.9\% of vehicles on average appear for less than 5 seconds in less than 4 cameras.

% !TeX root = main.tex

\begin{figure}
    \centering
    \includegraphics[width=0.48\textwidth]{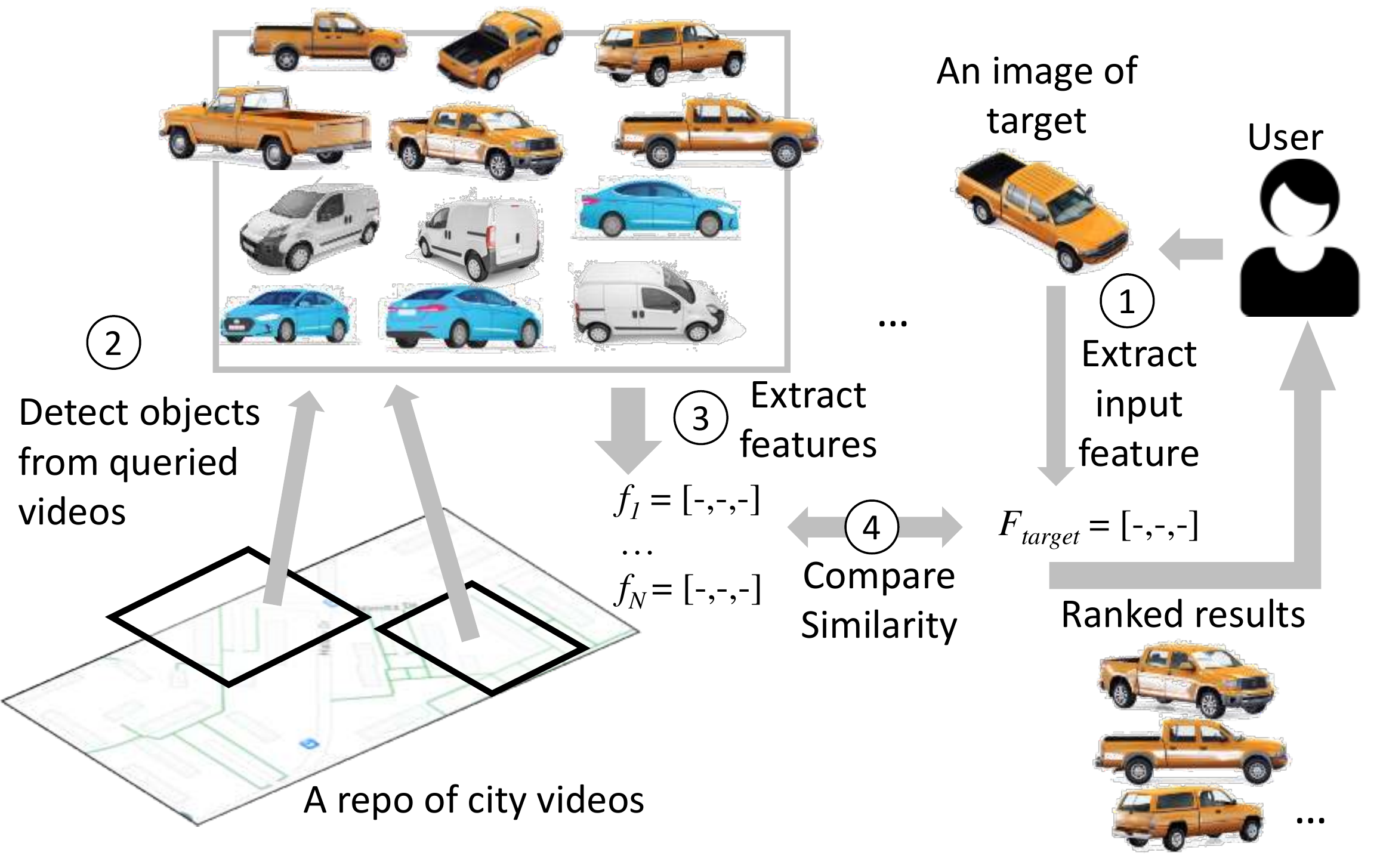}
    \caption{\textbf{The classic pipeline for object ReID}, formulated as image retrieval}
    \label{fig:pipeline}
\end{figure}

\noindent
\textbf{The common pipeline structure for ReID} is shown in Figure~\ref{fig:pipeline}: 
\Encircle{1} given an input image of target object X, 
the pipeline extracts its feature, e.g., using ResNet-152~\cite{resnet} to extract a 1024-dimension vector~\cite{tan2019cvpr, huang2019cvpr}; 
\Encircle{2} from the queried videos, the pipeline detects all bounding boxes belonging to the same \textit{class} as X, e.g., using YOLO~\cite{yolo}; 
\Encircle{3} the pipeline extracts features of all detected bounding boxes; 
\Encircle{4} it calculates pairwise similarities between X and the bounding boxes. 
The similarity is often measured as feature distance~\cite{l2norm}, where a shorter distance suggests a higher similarity between X and a bounding box. 
Of the four stages, stage 2 and 3 are most expensive. 
For instance, calculating feature distances in stage 4 is three orders of magnitude faster than extracting the features in stage 3. 
The cost of stage 2 and 3 further grows with the amount of videos. 
This pipeline structure is widely used, e.g., by almost all participants in popular vehicle ReID challenges~\cite{huang2019cvpr, tan2019cvpr, lv2019cvpr}.
Proliferating ReID algorithms call for a practical ReID system. 
Our driving use case is vehicle ReID, where personal identifiable information such as license plates are intentionally removed for privacy~\cite{cityflow}. 
Vehicle ReID is considered one of the most important ReID problems~\cite{aicity}. 
%There are mature use cases, 
% e.g., crime investigation and traffic planning, 
% commercial offerings~\cite{verkada} \Note{cite a few reID startups}, 
%(e.g., hard drives designed for surveillance~\cite{surveillancestorage}), 
%annual competitions~\cite{aicity}, 
%open competitions~\cite{aicity}, and public datasets~\cite{cityflow}. 
The techniques are likely transferable to other object classes. 

%\paragraph{Challenge 1: Unreliable bounding boxes}
\paragraph{Challenge 1: Limitations of modern ReID algorithms}
By its definition in computer vision, ReID focuses on differentiating numerous objects of the \textit{same class}, e.g., cars. 
In real-world videos, however, 
many objects of the same class exhibit minor visual differences; 
yet bounding boxes of the same object  -- captured by the same or different cameras -- may appear quite different.
% their corresponding feature vectors often show a high variation. 
As we will show in in Section~\ref{sec:motiv}, 
even sophisticated feature extractors
% bounding boxes of \textit{different} objects could show higher similarity than bounding boxes of the same object. 
%the similarity between bounding boxes of \textit{different} objects could be higher than that of bounding boxes of the same object. 
may deem bounding boxes of \textit{different} objects more similar than bounding boxes of the \textit{same} object. 
% ---- below could be problematic --- 
% For this reason, even equipped with multiple neural networks, modern ReID pipelines achieve a mean average precision %(mAP) of 0.7--0.8 on vehicles~\cite{tan2019cvpr,cityflow}. 
% That is, a substantial amount of bounding boxes returned by these pipelines are false positives. 

%This means that reID will 20-30\% false positives and XX\% false negatives on vehicle instances.  \Note{complete this}

% "American street network intersection densities typically range from as little as 60 intersections per square mile (one example is the street network in downtown Salt Lake City) to more than 500 (such as the network in downtown Portland, OR)."

\paragraph{Challenge 2: Many cameras, large videos} 
City cameras can be numerous. 
With 2-5 cameras per intersection \cite{cityflow,visionzero} and 60--500 intersections per square mile in urban areas~\cite{streetnetworks}, a query covering a few square miles would need to process a few hundred, if not a few thousand, cameras. 
%\Note{XXX be careful with numbers. need to state FPS, how many bounding boxes, etc}
Furthermore, modern ReID pipelines have an insatiable need for resources.
For example, Titan V, a $\sim$\$3,000 modern GPU, runs YOLO~\cite{yolov2, yolov2-web} for detecting bounding boxes at only 40 FPS.
The GPU running ResNet-152 extracts $\sim$80 features per second. 
% We estimate the GPU will run for $\sim$10 days for processing all bounding boxes in a month of videos from one camera. 
To process city videos from one square mile in a day, 
we estimate at least several hundred GPU hours are needed. 
This cost quickly becomes prohibitive as camera deployment and query scope expands. 
Resorting to cheap vision algorithms, e.g. smaller neural networks or SIFT, is unlikely to help: 
they are much more susceptible to subtle visual differences and disturbance, 
making ReID results even less usable. 
%\Note{why cite this}~\cite{dukemtmc4reid,rexcam}

% The family of ResNet~\cite{resnet}, typically used to extract object features~\cite{rexcam,tan2019cvpr}, processes images at most a few hundreds images per second~\cite{resnetspeed}.

\paragraph{Principles}
While prior research formulates ReID as image retrieval queries, i.e., to find every bounding box of a target object X~\cite{rexcam, jain18arxiv, huang2019cvpr, tan2019cvpr}, 
we treat ReID as \textit{spatiotemporal} queries, which search for what users care about: the times and locations in which object X appeared. 
%By focusing on time and locations rather than each bounding box, 
This gives opportunities to overcome the accuracy limitation on individual bounding boxes, and to quickly emit times and locations before processing all bounding boxes in the queried videos. 

%We the challenge of many cameras 
We address the multitude of city cameras by renewing a wisdom in video analytics: 
resource/quality tradeoffs~\cite{videostorm,noscope,focus,vstore,chameleon}. 
Prior video systems often target fewer cameras, 
making such tradeoffs \textit{within} a video stream, e.g., by tuning frame resolutions, rates, and cropping factors.
On city-scale videos, however, 
processing more cameras is almost always favorable than processing more pixels from each camera.
% it is almost always beneficial to prioritize camera quantities over video quality. 
% our insight is to prioritize camera quantities over video quality. 
% ``better to know a bit from many cameras than knowing much from a few cameras''. \Note{camera quantities over video quality?}
This is because: 
(1) cameras in different locations provide extensive spatial coverage; 
(2) cameras in the same location provide complementary viewpoints. 
Both factors benefit ReID queries more than video quality. 
To this end, we prioritize increasing camera coverage over increasing video quality, e.g., frame rate and resolution. 

We make minimum, qualitative assumptions on camera deployment.
Quantitative deployment knowledge, e.g., camera orientations and correlations, used to enable optimizations within smaller camera networks~\cite{rexcam}. 
For emerging city-scale cameras, however,
it is unclear if there exists a generic, quantitative deployment model. 
Minimizing assumptions allow a generic system design, 
which, as we will demonstrate, serves as the basis for deployment-specific optimizations.

\paragraph{\sys{}}
We present a ReID engine called \sys{}. 
Catering to spatiotemporal queries, 
\sys{} organizes all videos in a repository as \cube{}s, where a cell <$L$, $T$> contains video clips captured by all cameras near a geo-location $L$ during a time period $T$. 
%To query for a target object X, a user submits an image of X; 
\sys{} answers a query for target object X with a short list of \cubes{}, ranked by their promises of containing X; 
each returned cell is accompanied by video clips, with annotations of the likely bounding boxes of X. 
As executing a query, \sys{} keeps updating the rank based on new results from video processing. 
The user reviews returned cells and makes the final decision. 

We design and evaluate \sys{} as a recall-oriented system~\cite{recall-oriented}: 
it seeks to find all positive cells (which are rare) and rank them to the top.
As such, \sys{} minimizes human efforts in analyzing videos; 
it does not seek to replace humans, whose knowledge cannot (yet) be substituted by algorithms on real-world videos.
This goal is shared by existing recall-oriented systems, e.g., for legal documents or patents search~\cite{rank,choppy}, where final decisions from humans are indispensable. 

%The users terminate the query until satisfaction or the return diminishes. 
%Rather than returning final rankings in one shot, \sys{} presents rankings soon after a query starts; it keeps refining the rankings as it processes more video footage. 

\begin{comment}
% -- not very useful ----
    For instance, \Note{compress below} a police officer quickly spot the identified perpetrator in the neighborhood covered by the camera deployments, the information he/she needs is typically where, i.e., the cameras that captures, and when, i.e., the time that captures the target vehicle, rather than the retrieval vehicle images~\cite{rexcam}.
    (designed with user in loop in mind. continuous refining of ranking of results; instead of returning in one shot) This resembles a recommendation system, which returns a ranked list and the user will make final pick. 
\end{comment}

%\textbf{Key idea: Clustering of (unreliable) bounding boxes}
%\paragraph{Key design 1: clustering (unreliable) object features}
\paragraph{Key design 1: clustering fuzzy bounding boxes}
\noindent From a set of bounding boxes, 
how should \sys{} assess occurrences of a target object with confidence? 
Our insight is: how is an object perceived by a camera is 
heavily impacted by (1) the camera's posture, including position and orientation; 
(2) transient disturbance, such as occlusion and background clutter. 
These impacts sometimes overshadow the object's characteristics, e.g.,  shape and color. 

To counter the two impacts, \sys{} \textit{matches} the origin camera's posture: 
it samples diverse co-located cameras in hope of finding ones with postures similar to the input. 
%\sys{} seeks a camera that has a posture close to the one producing the input image. 
\sys{} \textit{mitigates} the disturbance:
it clusters similar bounding boxes captured by a camera during a period of time. 
Each resultant cluster thus represents the camera's general ``impression'' of a distinct object. 
\sys{} estimates the occurrence of X based on the similarity between the input image and distinct objects as represented by clusters. 
Clustering has been a classic algorithm in data processing~\cite{kmeanspaper,kmeanspaper2,hc} especially in vision~\cite{focus}; 
\sys{} is novel in applying it to ReID, deriving robust query answers from fuzzy bounding box features.
Clustering suits our principle of prioritizing camera coverage, 
as it tolerates low frame rate on each camera. 

\paragraph{Key design 2: Incremental search in \cube{}s} 
how to search in numerous \cube{}s and quickly find target objects?
% The key is to reconcile the needs for skipping redundant video contents and for exploiting diverse camera postures. 
The key is the camera sampling strategy: 
to avoid redundant video contents as much as possible while exploiting diverse camera postures as needed. 
%To this end, % \sys{} search all cells in the order of their promises of containing the target object. 
\sys{} navigates its resource spending towards cells where new discovery of target occurrences is most likely. 
% where any previously unknown occurrence of the target is most likely.
To do this, \sys{} starts a query with a minimum number of cameras to quickly estimate promises of all cells; 
it processes videos from additional cameras for undecided cells; 
it iteratively assesses cell promises and re-ranks all cells for subsequent search. 

We implement \sys{} and evaluate it on a video dataset of 25 hours of videos from 25 cameras. 
On 70 queries with different target objects, 
\sys{} delivers a high recall accuracy of 0.87 on average; 
it reaches high an accuracy goal of 0.99 in 108.5 seconds on average, at the speed of 830$\times$ of video realtime. 
Compared to alternative designs, \sys{} reduces query delays by up to 6.5$\times$. 
% and higher accuracy on 20\% of the queries. \Note{improve}
We further evaluate deployment-specific optimizations.

\paragraph{Contributions}
We made the following contributions.
\begin{myitemize}
\item Towards a practical ReID system,  % the limited algorithm accuracy and large numbers. 
we advocate a new approach: focusing on finding relevant \cubes{} rather than individual object instances. 

\item We present to cluster fuzzy bounding boxes as approximations of distinct objects, which effectively overcomes limitation in ReID algorithm accuracy. 

\item We present incremental search in cells. This minimizes redundant processing while exploiting diverse camera viewpoints, which reduces the ReID compute cost. 

\item We report \sys{}, a ReID system that works on large video repositories.

\end{myitemize}

\paragraph{Ethical considerations} 
In this study: all visual data used is from the public domain; no information traceable to human
individuals is collected or analyzed.

% In our design, we eschew any license plate related recognitions, and the dataset~\cite{cityflow} used in our evaluation has the license plates intentionally blocked to protect user private data and avoid privacy leakage.
% In addition, vehicle license plates does not always act as a solid measure in vehicle re-identification problems because license plate can be easily falsified in vision data.
 % 2 pgs
    % !TeX root = main.tex

\section{Motivations}
\label{sec:motiv}
% Based on our observations on camera deployments, video data, and CV operators, many assumptions made towards object re-identification have and will depart from prior work~\cite{rexcam,caeser} in the upcoming new era of smart cities, \Note{too strong. likely backfire} and there is an urgent need to reinvent a new query processing engine.

\subsection{System model}

% !TeX root = main.tex

\begin{figure}
    \centering
    \includegraphics[width=0.45\textwidth]{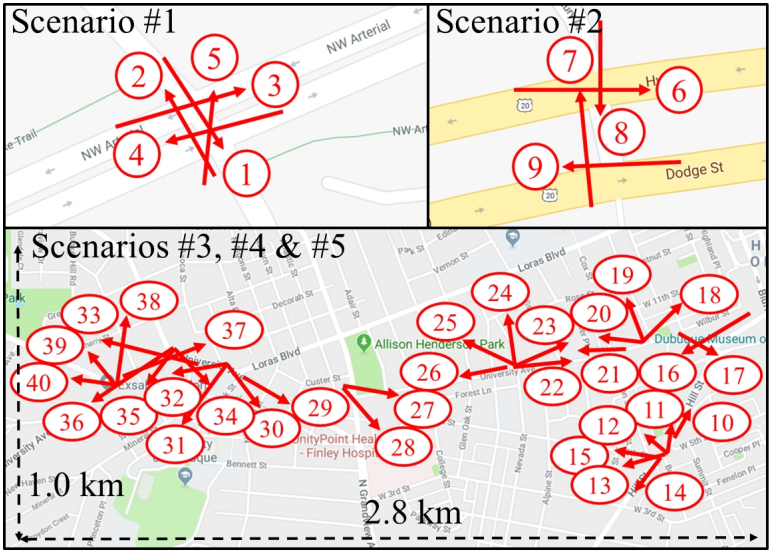}
    \caption{\textbf{An example of city camera deployment that motivates our design.} 
    Each red arrow: a camera's location (the arrow tail) and orientation (the arrow direction);
    Numbers: camera IDs.
    Both data and figure from CityFlow~\cite{cityflow}. 
    }
    \label{fig:cityflow}
\end{figure}

\paragraph{Queries \& videos}
We target retrospective queries: 
at the query time, all videos are already stored in a central repository. 
We assume a large repository of videos from geo-distributed cameras. 
Preprocessing at ingestion, i.e., as videos are being captured, is optional, as permitted by compute resource. 
At ingestion time, the system knows the object classes that may be queried, e.g., cars, 
but not the input images of queries. 
% However, ingestion preprocessing does not know about future queries and is agnostic to specific queries.

A query includes an input image of the target object X and the scope of videos to be queried; 
the query does not carry any metadata, e.g., a timestamp or the origin camera that produced the input image.
Following the norm in ReID research~\cite{sun18eccv,zhong17cvpr,fu19aaai,tan2019cvpr,huang2019cvpr,fu19iccv}, we do not assume the video repository contains the input image; 
we do not assume any other images from the origin camera is available. 

% it may \textit{not} contain any video from the origin camera. 

% It is uneconomical/undesirable to process all the videos at ingestion:  detecting bounding boxes and extracting features from all bounding boxes \Note{the last point questionable}. 

\paragraph{Cameras}
% We motivate our design with the following observations. 
We make minimum, qualitative assumptions on camera deployment. 
The deployment covers multiple geo-locations. 
At each location, multiple cameras are co-located as a \textit{geo-group}. 
The query system knows which cameras are co-located, i.e., belonging to the same geo-group.
Of the same geo-group and during a short period of time, e.g., tens of seconds, cameras are likely (although not necessarily) to capture similar sets of objects from different viewpoints. 
A deployment example is shown in Figure~\ref{fig:cityflow}. 
% These assumptions hold for popular city video datasets~\cite{cityflow}. 

We do not assume that the query system knows quantitative camera postures and quantitative correlations across camera geo-groups (e.g., ``one object appearing in geo-group A has 50\% chance to reappear in geo-group B within the next 10 minutes''). 
There is no sufficient evidence showing such knowledge is standard in city camera deployment.
As such, we design a generic system without such knowledge, 
and evaluate how the resultant system can be augmented in case some knowledge becomes available (Section~\ref{sec:eval}). 

\subsection{Challenge 1: Algorithm limitations}
\label{assumptions:cv}

% \paragraph{Clustering is more robust in ranking \textit{trips}}
% To derive vehicle similarities, the euclidean distance~\cite{l2norm} of features vectors between the a pair of vehicle images are measured.
% We have observed the following behavior from DNNs in vehicle re-identification tasks.

% !TeX root = main.tex

\begin{figure}
\centering
\includegraphics[width=.99\linewidth]{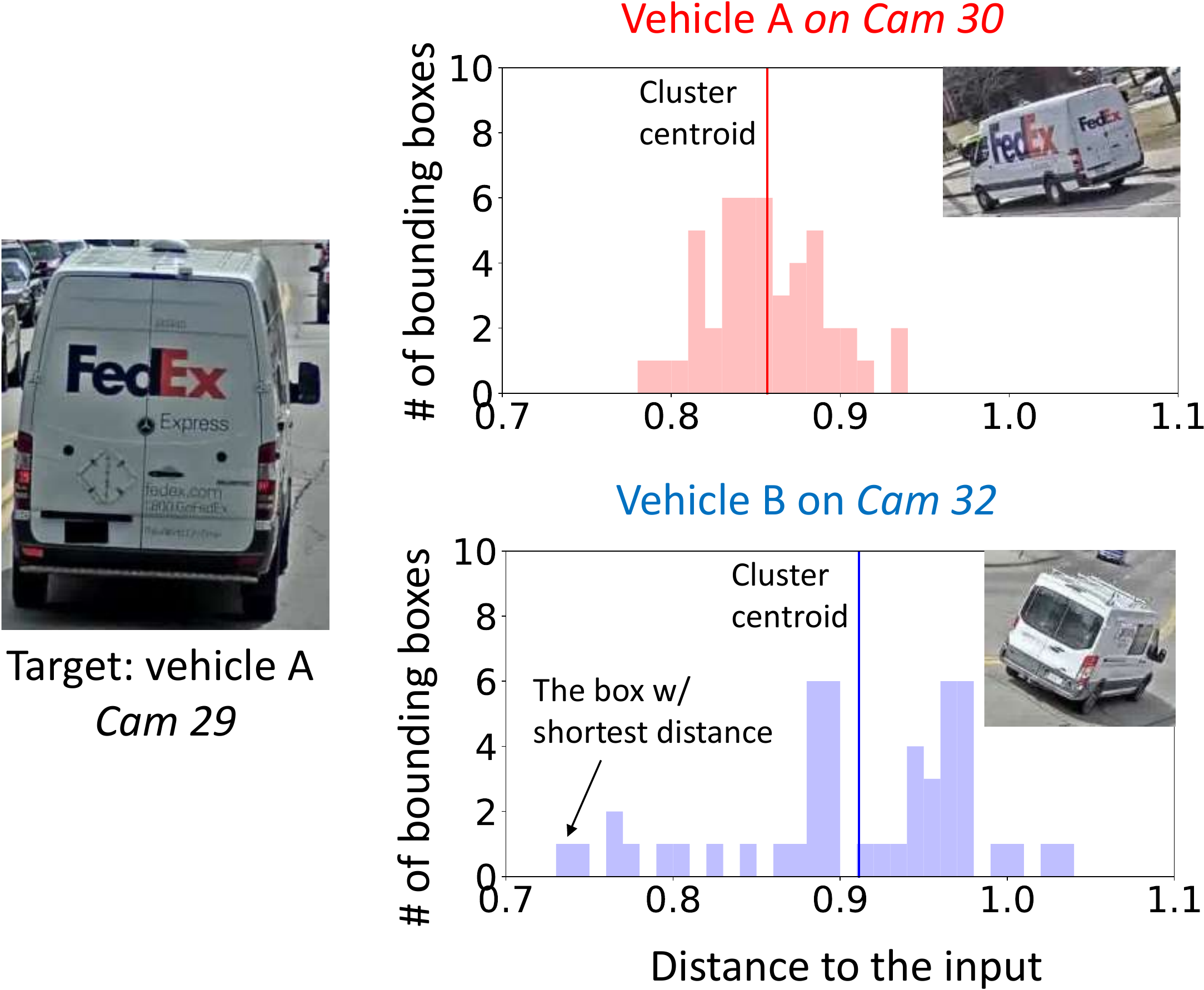}
\caption{\textbf{Examples of fuzzy bounding boxes}. 
(Left) an image of vehicle A, whose feature is the input.
(Top) a histogram of distances between the input and other features of A.
(Bottom) a histogram of distances between the input and features of B, a confusing vehicle.
All features are 1$\times$1024 vectors extracted by ResNet-152. 
Euclidean distances with L-2 norm~\cite{l2norm} are used.
Video clips: 4.7/4.9 sec for vehicle A/B from CityFlow~\cite{cityflow}}
\label{fig:nn}
\end{figure}

% ($1\times1024$ vector) euclidean distances (L-2 norm~\cite{l2norm}) distributions between target vehicle A (camera 29), and the same vehicle A(camera 30, 4.7s) and a different vehicle B (camera 32, 4.9s); NN: ResNet. Problem: a very diff viewpoint works better. Maybe omit the actual image? XXX We show examples from CityFlow~\cite{cityflow}, a suite of benchmark videos collected from real-world city cameras. XXX}

% We use CityFlow~\cite{cityflow} to illustrate the limitations of reID algorithms.
% CityFlow is a video dataset captured by real-world city cameras and city reID video 

% We show examples from CityFlow~\cite{cityflow}, a suite of benchmark videos collected from real-world city cameras. 

\paragraph{Observation: fuzzy bounding boxes}
Figure~\ref{fig:nn} compares the features of a target vehicle A and a confusing vehicle B. 
The features are extracted by ResNet-152, a state-of-the-art neural network.
Given an image of vehicle A: 
(1) ResNet-152 deems 10\% of B's bounding boxes exhibit shorter feature distances than A's, hence are more similar to the input image, as compared to A's other bounding boxes; 
(2) the bounding box closest to the input image is from the confusing vehicle B but not A; 
(3) Bounding boxes of the same vehicle show a high variation, as reflected by the wide range of the feature distances. 
The above example of fuzzy bounding boxes is not isolated: 
they are the major hurdle for ReID accuracy, responsible for an average of 0.65 loss of accuracy in 20\% of queries executed by a baseline design (\sect{eval}).

\begin{comment}
\paragraph{Key opportunity: Clustering feature vectors}
When determining the ranking of feature vectors coming from different vehicles, \sys{} first clusters the feature vectors into several groups and typically, feature vectors coming from the same vehicle are clustered into the same cluster dues to its internal similarity.
By comparing $f_{target}$ with the cluster centroids, as shown in Figure xxx, we alleviates the impact of those outliers from false positive vehicles that have a relatively small distance with $f_{target}$, as the centroid could represents an overall feature of this vehicle, thus derives better ranking accuracy when considering vehicle \textit{trips}.
\noindent Well-known clustering methods like K-means clustering shows promises.
We conducted K-means clustering on each of video clips in CityFlow~\cite{cityflow}, and the average clustering accuracy reached as high as 92.87\%~\cite{coclust}.

% Other clustering method, e.g., hierarchical clustering, still suffers from the entanglements because those algorithms are still starting from merging feature vectors based-on individual similarities.

%\Note{the following sentence unclear}
% We have observed that the re-identification inference accuracy can be improved by considering vehicle similarities as entirety rather than individuals, typically adopted by prior work~\cite{rexcam}.
% As shown in Figure~\ref{fig:nn}(b), we have observed a prevalent presence of outliers coming from a false vehicle: the features vectors from two different vehicles around 0.8 are heavily entangled.
\end{comment}

\paragraph{The causes}
Why fuzzy bounding boxes? 
We explain the root cause with a simple formula: 

\begin{equation*}
%\label{eq:cam}
%	Cam'(f) = Cam(f) + \mathscr{N}
Cam'(X) = Cam(X) + \mathscr{N}
\end{equation*}

\noindent
This formula describes 
how $Cam'(X)$, a camera's actual observation of an object, is formed. 
% how F(X,C), an object X's feature as perceived by a camera C. In this formula: 
$X$ is the object's inherent characteristics, e.g., its color, shape, skeleton, and key points.
Two factors prevent a ReID system from directly learning $X$ and matching it to the input. 
First, a camera's posture modulates $X$ as Cam($X$), i.e., the camera's \textit{ideal} observation on X.  
% avoid saying angle and distance, 
Second, the ideal observation is susceptible to transient disturbances $\mathscr{N}$, e.g., changes in resolution and viewpoint as objects move, 
background clutter, and occlusion. 
The impacts of camera postures and transient disturbances are strong, sometimes even stronger than the impact of inherent characteristics $X$. 
We have observed a different camera viewpoint of the same object resulting in 3$\times$ difference in feature distances. 
% For instance, we have observed that given an input image, by calculating the pairwise feature distances between the same vehicle from a similar viewpoint and a different viewpoint, those distances can have up to 3$\times$ differences.
Hence, classic ReID pipelines that aim at labeling each bounding box are in fact deciding on $Cam'(X)$, 
which encodes the camera posture and also the disturbances. 
The pipelines cannot achieve high accuracy because it is difficult to model $Cam()$ and $\mathscr{N}$ and eliminate their impacts accordingly.

\subsection{Challenge 2: Numerous cameras \& videos}
\paragraph{Colossal data volume}
A city camera generates more than 6 GBs of videos daily (720P at 1FPS).  
Estimated from recent reports on camera deployment in northern American cities~\cite{visionzero,cityflow,streetnetworks}, 
the number of city cameras per square mile ranges from a few hundred to a few thousand. 
A ReID query covering only a few square miles and one day of videos will have to consume PBs of videos.

\paragraph{Expensive pipelines}
Extensive work has been proposed to use neural networks (NNs) for ReID, 
advancing the accuracy steadily~\cite{paperwithcode} on public datasets~\cite{dukemtmc4reid, market1501}. 
% boosting the accuracy by more than 5$\times$~\cite{paperwithcode} on public datasets~\cite{dukemtmc4reid, market1501}. \Note{5x too much? so we still have low accuracy? check the number again} at the expense of high compute cost.
For instance, recent pipelines cascade multiple NNs, each detecting a separate set of vehicle attributes, e.g., orientations and roof types. 
The additional NNs are reported to improve accuracy (mAP) by 10\% with up to 7$\times$ overhead~\cite{tan2019cvpr,huang18cvpr}. 
We estimate that they can run no more than 15 FPS on a modern GPU. 

\paragraph{Would cheaper features help?}
Cheaper NNs and vision primitives are unlikely remedies. 
The former were used by many \textit{object detection} systems to provide a middle ground between high accuracy and low cost~\cite{noscope,focus,elf,filterforward}.  
On the much harder ReID tasks, however, modern NNs simply do not offer surplus accuracy for systems to trade off. 
We have tested RGB and SIFT, two cheap vision primitives for extracting object features. 
RGB features are highly volatile to lighting conditions and background clutter; 
SIFT yields poorer features compared to modern NNs, while not running significantly faster than the latter. 
Prepending these primitives to a ReID pipeline are likely to hurt performance.

\begin{comment}
Cheap early filters are even fuzzier and are thus not designated for reID. 
For instance, we have tested color filters like RGB histogram(cite): although it can process thousands of images per second, the RGB values are highly subject to lighting conditions and background; 
Local feature matching algorithms like, SIFT, SURF, ORB (cite) do not exhibit significant cost reduction but are much less robust compared to advanced DNNs on reID tasks.
Cheap NNs with only a few layers, .e.g., a shallower AlexNet(cite), though widely used in prior work~\cite{noscope, focus} (cite more) with up to three orders of magnitude speedup compared to full-fledged NNs, e.g., ResNet, are only proved to serve well on easy object recognition tasks while not on reID tasks.
\end{comment}

% model ensembles 
% There also exists attributes classification, e.g., vehicle type, roof rack, etc., but only have little improvements~\cite{tan2019cvpr}.
% they also concatenate the orientation feature by detecting positions of vehicle key points~\cite{huang18cvpr,tan2019cvpr}.

\subsection{Why is prior work inadequate}

\noindent
\textbf{Computer vision research}
typically treats ReID as image retrieval~\cite{reid-survey, tan2019cvpr,huang18cvpr,lv2019cvpr}. 
Aiming at finding all bounding boxes of a target object, 
computer vision studies typically focus on improving accuracy without considering query speed or efficiency much. 
Yet, retrieving every bounding box would miss opportunities, as we will show, 
that can provide useful spatiotemporal answers with much lower delays.

\noindent
\textbf{Existing ReID systems}
often consider smaller camera deployment and are evaluated on such datasets, 
e.g., 8 cameras over a university campus~\cite{dukemtmc4reid}.  
Many core designs depend on deployment-specific knowledge. 
% Their core design assumptions do not necessarily hold for city scale cameras. 
For instance, ReXCam~\cite{rexcam} searches in cameras among which spatial correlations are both strong and known.
Given an input image captured by a \textit{known} camera in the network, 
the system exploits camera correlation to find all images of the object. 
ViTrack~\cite{vitrack} models and then predicts object trajectories in answering ReID queries. 
% ViTrack~\cite{vitrack} models camera correlations to predict the object's entire moving trajectory in missed time ranges.
% in the same trip traveling through the camera network. 
However, some assumptions (e.g., camera correlations) do not necessarily hold at the city scale; 
some others (e.g., a known origin camera) restrict use cases and are incompatible with the norm in ReID research~\cite{sun18eccv,zhong17cvpr,fu19aaai,tan2019cvpr,huang2019cvpr,fu19iccv}.
Without such strong assumptions, we intend our base design to be generic, 
and can nevertheless optimize queries with additional information as they become available (\sect{eval}). 

\noindent
\textbf{Spatiotemporal databases} are designed for managing object trajectories, e.g., airplane movements and human movement, and answering queries on them~\cite{geomesa,erwig99geoinformatica, abraham99geoinformatica,dbsurvey,porkaew01stdb}. 
They ingest \textit{structured} data, e.g., sequences of time-location tuples as from GPS; 
they cannot ingest unstructured video data and recognizes object occurrences as \sys{} does. 
The output of our system can be the source of a spatiotemporal database.

\begin{comment}
However, those work does not support querying unstructured video data as no ground truth labels are created.
Besides, different from recovering the object's trajectory from dense data points, \sys{} is built for sparse camera groups.
Unlike low-power localization devices like GPS, it is uneconomical/infeasible to deploy cameras everywhere in the city, as human are usually interested in specific locations, e.g., intersections.
\end{comment}

 % 1.5 pgs
    % !TeX root = main.tex

\section{\sys{}}
\label{sec:overview}

%\Note{copied here, to merge}
%At query time, given only a snapshot of the target vehicle, the query engine extracts its feature vector $f_{target}$ and searches for its \textit{trips}, i.e., locations and time ranges in which the target vehicle appears, in the video footage captured.
% in the video footage from all cameras, in fixed-length time windows, e.g., 30s, from each camera.
% It then returns a ranking of the video clips indicating the spatial information, i.e., the camera that captured the target vehicle and temporal information, i.e., the time range that the vehicle appears on each camera, to the user.

\paragraph{Video ingestion}
At ingestion time, \sys{} optionally pre-processes videos from a small number of cameras, as permitted by available compute resource. 
The pre-processing detects objects of interesting classes (e.g., cars) and extracts their features; it is agnostic to specific queries.
The pre-processing is elastic; \sys{} will run unfinished pre-processing at the beginning of a query's execution. 

\sys{} runs profiling as it ingests videos, a common practice of video systems~\cite{focus,rexcam,elf}.
It periodically samples videos from each camera to train several parameters used in clustering bounding boxes and  determining camera sampling order. %, and filtering query results. 
We will discuss these parameters in detail in Section~\ref{sec:clustering} and \ref{sec:query}. 
The profiling is light, processing 30 seconds of every 1-hour video and taking less than 10 seconds on a modern GPU. 

\begin{figure}
\centering
    \includegraphics[width=0.49\textwidth]{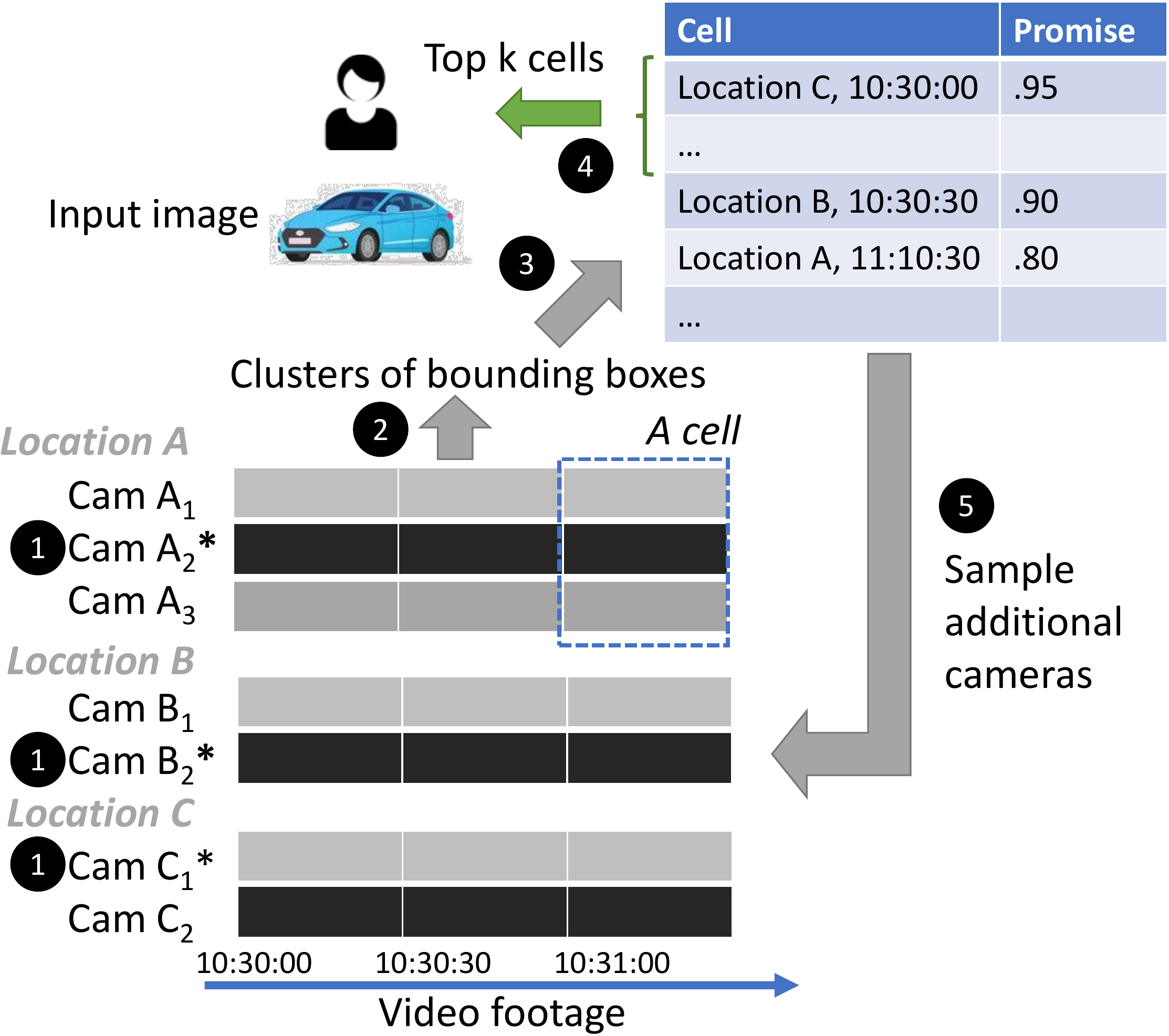}
    \caption{\textbf{An overview of \sys{}}. * = a starter camera}
\label{fig:overview}
\end{figure}

\paragraph{Executing a query}
\sys{} organizes all the queried video footage by cells, 
each consisting of video clips captured near a location during a fixed period, 
as shown in Figure~\ref{fig:overview}.

To execute a query, \sys{} searches in all cells iteratively; 
it adds cameras to each cell for processing in an incremental fashion. 
It starts by sampling from all cells in the query scope. 
The initial sampling is brief, as it only processes a small fraction of video footage in each cell -- from selected cameras (``starter cameras'') \encircle{1}. 
% In this way, \sys{} quickly samples from all cells at low cost. \encircle{2}
From the sampled video of a cell, \sys{} detects distinct objects out of fuzzy bounding boxes; 
it does so by clustering features of bounding boxes by similarity. 
\sys{} treats each resultant cluster representing a distinct object, 
where the cluster's centroid is an approximation of the object's feature \encircle{2}. 
\sys{} ranks all the cells by their promises, estimated from similarity between their enclosed objects and the input image \encircle{3}. 
\sys{} emits the ranked cells as query results to the user, who reviews the top ones \encircle{4}. 
\sys{} selects additional cameras for processing and uses the results to update the cell list continuously \encircle{5}. 
A query is terminated by the user manually (e.g., when she is satisfied with results) or when \sys{} finishes processing videos in all queried cells. 

\paragraph{Limitations}
\sys{} inherits the statistical nature of its underpinning ReID algorithms, notably the neural networks. 
While \sys{} empirically shows high confidence in its query results, 
e.g., it finds all true cells in more than 70\% of queries (\S\ref{sec:eval}), 
% TODO: this is actually top 5,
it, however, cannot provide sound guarantee to do so. 
Similarly, although \sys{}'s accuracy often quickly converges during query execution, 
there is no guarantee on the convergence rate, e.g.,  processing 50\% of videos to reach accuracy of 0.75. 
The hope is that users review the top k cells for true results; 
they entrust \sys{} on the remaining, unreviewed videos, being comfortable with the level of confidence that \sys{} provides. 
However, in case they want to be absolutely certain that no true cells are left out, 
they would need to inspect all the videos.

% moved from ``discussion'', can be useful

\begin{comment}
\paragraph{Limitations}
\sys{} never adopts any license plate recognition technique, so that for two vehicles with the same make, model, and color, DNNs are not be able to effectively differentiate between them.
Any state-of-the-art solutions on object re-identification are limited in distinguishing the a pair of object that has the exact same appearances, e.g., vehicles having the same make, model, and color.
\sys{} reduces the chances of by treating object re-identification queries as spatio-temporal queries, and vehicles inside a small time range, e.g., 30s, are usually less likely to be exactly the same.
\Note{What if clustering makes mistakes? Can go to limitation discussion}

\paragraph{More input images}
\sys{} could benefit from multiple input images that capture the vehicle at different viewports, as there will be a higher chance that the input images match the camera viewport.
\end{comment}
 % .5 pg
    % !TeX root = main.tex

%\section{Clustering Object Features}
\section{Clustering bounding boxes}
\label{sec:clustering}

A core mechanism of \sys{} is to recognize distinct objects from bounding boxes and compare the recognized objects to the input image. 
It addresses two concerns: 

\begin{myitemize}
\item Working around fuzzy features of bounding boxes resulted from the limitations of ReID algorithms. 
\item Tolerating low frame rate, which allows \sys{} to sample more cameras, one of our principles in \sect{intro}. 
\end{myitemize}

\paragraph{Observations on transient disturbance}
% How to match an input image to numerous unreliable bounding boxes, each encoding a camera's specific posture and transient disturbance?
% By sampling from extensive cameras, \sys{} already exploits cameras diversity, in hope of that postures of some cameras are close to the origin camera. 
How to match an input image to numerous fuzzy bounding boxes, each encoding impacts of transient disturbance?
The disturbance is time-varying and its impacts can be either graduate or sudden. 
For example, as a vehicle travels through a camera's field of view, 
its bounding box may resize; 
its view angle may change; 
occasionally, it may be occluded by a light pole; 
its background may be intruded by another vehicle. 
% Graduate impacts, e.g., shrinking a bounding box, often distort bounding box features,  while sudden impacts will disrupt the features and make them outliers. 
% accounts for up to more than $2\times$ extra pairwise distance. 

% The object resolution depends on the size of the bounding box detected, and we have observed that having a similar resolution with the input image usually has a higher pairwise similarity, as DNNs always take fixed input size and requires resizing before extracting the object feature.
% As mentioned in Section~\ref{sec:motiv} and Figure~\ref{fig:overview}, we have observed that a very different viewport could double the feature distances.

\paragraph{Key idea: clustering similar bounding boxes}
Disturbance to bounding box features is difficult to model and eliminate in general.  
Yet, if we consider similar bounding boxes in consecutive video frames, 
their distorted features due to graduate impacts are likely to smooth out, 
and the outlier features due to sudden impacts can be removed from consideration. 

To this end, \sys{} clusters object features based on their similarities. 
The similarities, for instance, can be measured by Euclidean distances across 1024-dimension feature vectors. 
As a result, each cluster represents a camera's \textit{general impression} on a distinct object during a given time window. 
The cluster's centroid is an approximation of the camera's ideal observation of the object. 
%\notettx{but input image can also suffer from disturbances?}
%

\sys{}'s use of clustering is novel,
in that it overcomes the accuracy limitations on individual bounding boxes. 
Notably, it differs from prior video systems~\cite{focus} that cluster objects for efficiency, e.g. to avoid  processing similar objects in a cluster. 

Figure~\ref{fig:nn} showcases why clustering is useful. 
% the most similar image comes from vehicle B (the leftmost blue column) that has the smallest pairwise feature distance of 0.74.
Recall that this example shows the difficulty in comparing an input image to individual bounding boxes (Section~\ref{sec:motiv}).
However, once we cluster the respective bounding boxes of the two vehicles, 
the centroids (distances as solid vertical lines) are much more robust indicators of object similarity, suggesting that the general impression of vehicle A is much closer to the input image. 
%the centroid (mean) feature distances (solid vertical lines) represent \textbf{a general ``impression''} of a vehicle, and it could \textbf{smooth out (cancels) the occasional disturbance (transient impact)} due to viewports, occlusions, etc. 

%\Note{not sure why we mention the following}
%Other cheaper vision operators, e.g., RGB histogram, SIFT, are even less reliable than NNs in feature extraction, thus cannot effectively distinguish among different vehicles. \notettx{can skip this}

In practice, we find simple clustering algorithms often suffice. 
\sys{} runs k-means clustering~\cite{kmeanspaper} within each \cube{}. 
By minimizing the sum of intra-cluster variances across all clusters, 
k-means thus effectively puts most visually similar objects in the same cluster. 
k-means guarantees convergence to local optimum and is known robust to outliers~\cite{kmeanspaper2}.
We also tested other popular clustering methods, e.g., hierarchical clustering~\cite{hc}. 
% , which merges the most similar pairs of clusters in multiple iterations. 
We find them less favorable than k-means, e.g., they often attribute bounding boxes of the same object to separate clusters. 
% Besides, when clustering hierarchically, the stop criterion is hard to decide, \Note{is it same as deciding a K?} and a slight difference \Note{of what?} may ends up with very different clustering results. 

\paragraph{Predicting the number of distinct objects ($k$)}
As a prerequisite for applying k-means in a video clip, \sys{} must specify $k$ as the number of distinct objects in that video. 
An accurate $k$ is crucial to the clustering outcome. 
% It significantly reduces misclassifications for visually similar vehicles that are caused by NN's internal limitations.
% However, the value of K cannot be known beforehand, and therefore we propose to train a simple linear regression model based-on historical data for K-prediction when conducting K-means clustering.
% The estimated K should not deviate from the true K because then the clustering result accuracy will be severely compromised.

\sys{} predicts $k$ based on a simple intuition: 
of a given video scene, the distinct vehicle number is correlated to the spatial density of bounding boxes. 
Therefore, \sys{} only needs twofold information to predict $k$: 
(1)  $x_1$, the number of bounding boxes detected in the video clip; 
(2) $x_2$, the number of frames that contain non-zero objects. 
Such information is already available from object detection, the ReID stage that precedes feature extraction described in Section~\ref{sec:intro}. 
From $x_1$ and $x_2$, \sys{} further derives three variables 
as different orders of the box/frame ratio: $x_3 = (x1/x2)^2$, $x_4 = (x1/x2)$, and $x_5 = (x2/x1)$. 

We formulate classic kernel ridge regression~\cite{ridge}:     
$k = \mathbf{ax}+b $. 
The model takes as an input $\mathbf{x} = [x_1, x_2, x_3, x_4, x_5]$ which consists of all aforementioned variables; its parameters are a vector $\mathbf{a}$ and a scalar $b$. 
We instantiate one model for all the cells, for which we train $a$ and $b$ offline in one shot on 30-second labeled videos from 25 cameras.

% Prior video analytics systems~\cite{noscope,focus,chameleon,vstore,videostorm} adopts frame sampling that only process video data on a small portion of frames.
%However, object tracking is not robust on low frame rates, a common knob tuned by prior systems~\cite{noscope,focus,chameleon,vstore,videostorm} to reduce query costs.

% !TeX root = main.tex

\begin{figure}
\centering
\includegraphics[width=.99\linewidth]{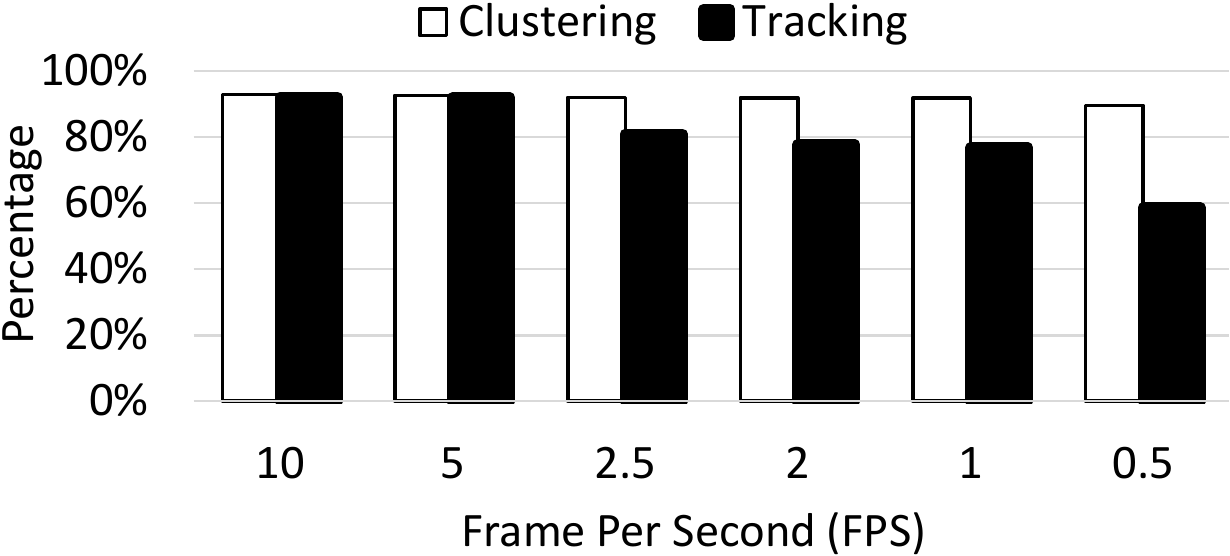}
\caption{\textbf{Clustering of bounding boxes tolerates low frame rates.}
Y-axis: the percentage of bounding boxes correctly attributed to respective objects. 
Object tracking implemented in OpenCV 3.4.4. Videos from CityFlow~\cite{cityflow}}
\label{fig:tracking}
\end{figure}

% XXX we use 1fps later. Can connect to this figure?? XXX should accuracy be percentage? be consistent. tacking - typo. Use columns instead of lines? XXX)

\paragraph{Tolerance of low frame rates}
k-means clustering is robust to low frame rates, suiting our design principle stated in \sect{intro}. 
As a comparison, we have investigated object tracking, another well-known approach to differentiating objects~\cite{zhu2018eccv, li2018cvpr}: 
first detecting individual bounding boxes on all frames; 
then linking bounding boxes across consecutive frames as distinct objects based on estimation of their motion trajectories. 
Yet, to estimate trajectory with good accuracy, object tracking demands a much higher frame rate than clustering. 
Figure~\ref{fig:tracking} shows an experiment: 
while both k-means and object tracking can identify distinct objects well with high frame rates, e.g., 10 FPS, 
as the frame rate drops to 2.5 FPS or lower, object tracking quickly loses accuracy to be barely useful ($\sim 0.6$ with 0.5 FPS). 
By contrast, k-means still maintains a high accuracy over 90\%. 
%k-means only sees a decrease of 3.16\% \Note{is this percentage, or percentage point}
%by reducing the frame rate from 10 fps to 0.5 fps, K-means clustering can still reach an accuracy of ~90\%
%~\cite{coclust} (\Note{what is this citation for}). 
Object tracking also suffers from other difficulties, e.g., differentiating multiple nearby objects following similar trajectories~\cite{motsurvey, mot}. 

Increasing frame rate for clustering, on the other hand, leads to a diminishing return: 
the accuracy improves by less than 3\% by increasing from 1 to 10 FPS. 
This supports our principle: prioritizing camera coverage over video quality.

% Tracking through Intersect Over Union (IOU)\notettx{cite}, i.e., the the overlapped region size of bounding boxes detected, between consecutive video frames, also does not work well under low frame rates.
% \Note{Need a subsection on challenges. (What are they?) Then transition into our key techniques}

% \paragraph{Viewport prediction are expensive}
% Although prior CV solutions added the viewport prediction.

 % 1 pg
    % !TeX root = main.tex

%\section{Searching in \cubes{}}
\section{Incremental search in \cubes{}}
\label{sec:query}

% \Note{the two metrics: promise and vote. What is per cell/camera? What is per cell? what is used for deliver result? for guide future search?  Need to clarify. Perhaps in Fig 7}

\begin{comment}
\Note{may remove}
To this end, the key is to sample cameras judiciously:
quickly covering as many geo-locations as possible;
for each location, processing as few cameras as possible (to minimize processing redundant contents), 
while processing as many as necessary (to exploit diverse camera postures). 
\end{comment}

\sys{}'s search mechanism addresses two design questions: 
(1) how likely does a cell contain the target object; 
(2) for which cells \sys{} should process additional cameras.
The former question determines the order of \sys{} processing undecided cells and the order of \sys{} presenting decided cells to users for review.
The latter question guides \sys{}'s search direction. 

\subsection{Assessing cell promises}
%\subsection{Assessing confidence in cells}
\sys{} estimates how likely a cell contains the target object by \textit{promise}. 
The rationale is that a cell shows high promise as long as any object in this cell is highly similar to the input image. 
Based on this rationale, 
we define the \textit{single-camera promise}, $p_{single}(R,\mathscr{C})$, as the promise \sys{} would perceive in cell $\mathscr{C}$
by only processing a video clip from a camera $R$:
it is \textit{reciprocal} to the smallest feature distance between the input and any centroid of object clusters from the video. 
That is, $p_{single}(R,\mathscr{C})=min(dist(X, o))$ where $ o \in objs$ and $X$ is the feature of target object. 
As \sys{} processes video clips from additional cameras for a cell $\mathscr{C}$,
it estimates the cell's overall promise, i.e., multi-camera promise, 
as the highest of single-camera promises of $\mathscr{C}$. 

The promise metric reflects our intuition: a cell appears more promising to \sys{} (with a higher multi-camera promise) as long as the cell has one strong champion camera (showing high single-camera promise) than having multiple weak supporters (medium single-camera promises). 

%the smallest  centroid feature distance ($d_{min}$) for centroid feature distance $d$ with $f_{target}$ by L2-norm across each clusters ($C_1 ... C_K$), the smallest distance ($d_{min}$) 
%(the minimum distance among all $d$s in each \cube{}). 

\subsection{Prioritizing cells in search}
A cell's promise reflects the single most similar object recognized in a cell. 
It, however, is inadequate for \sys{} to decide whether a cell is worth further exploring, i.e., to process more cameras for the cell. 
To do so, 
%\sys{} needs to track the \textit{accumulated} search efforts spent on the cell already and the \textit{accumulated} positive evidences 
\sys{} needs to track the \textit{accumulated} evidence discovered in the cell and the \textit{accumulated}
search efforts spent on the cell so far. 
% and how much opportunity is left for the cell. 
Neither is reflected in the cell's multi-camera promise.
For instance, if \sys{} only stops processing additional camera for cells with high enough promise, 
it may process too many cameras for cells where no single object is highly similar to the input. 

% \Note{thought: there should be some penalty on a cell's overall confidence, if we process a cam for a cell but do not see a gain. The rationale is that we checked one cam for that cell but didn't find things useful. The hope diminishes.}

To this end, \sys{} puts all the cells in three categories: 

\begin{myitemize}
    \item 
    The green cells: \sys{} has collected enough evidences -- though not necessarily all -- for them,
    and predicts them \textit{likely} to contain the target object. 
    % (although promises are not necessarily high). 
    Processing additional cameras is unlikely to change this assessment. 
     
    % is confident that these cells contain the target object; 
    
    \item 
    The red cells: \sys{} has collected enough evidences  
    and predicts them \textit{unlikely} to contain the target object.
    Processing additional cameras is unlikely to change this assessment. 
    
    % \sys{} is confident that these cells contain no target object. 
        
    \item 
    The gray cells: 
    the existing evidences are insufficient. 
    % \sys{} needs more evidences (i.e., a ``second opinion'' from another camera) to decide on these cells. 
    Processing one or a few cameras will likely turn the cells to red or green.     
    This is how a human analyst would make up mind on a suspicious cell -- by inspecting additional video footage from a different camera viewpoint. 
\end{myitemize}

\paragraph{Search plan}
All cells will begin in gray. 
\sys{} navigates its search from the gray cells (to resolve the undecided cells), 
to the green cells (to refine the order in which they will be presented to the user ), 
and then to the red cells (in the unlikely event of any true cells are left out). 
Based on new processing results in a cell, \sys{} updates its category accordingly as will be discussed below. 
\sys{} will exhaust processing cells in a category before moving to the next category. 
In each category, it always processes the cell that has the highest multi-camera promise. 

 % \Note{tie to Fig 7 which can be simplified. show the rank, show which portion is returned/visible to user, which part is for exploration}

\paragraph{Categorizing a cell with voting}
To determine the category for a cell $\mathscr{C}$, \sys{} runs a simple voting mechanism to incorporating observations of multiple cameras. 
The voting mimics how humans would make decision out of a set of expert opinions. 
\sys{} quantizes all single-camera promises with two thresholds, $P_{high}$ and $P_{low}$. 
To $\mathscr{C}$, a camera with promise $p$ casts a high-confidence vote with a weight of 1 if $p > P_{high}$;
it casts a medium-confidence vote with a weight of 1/k if $P_{high} < p < P_{high}$. 
The resultant vote count is more intuitive and tunable than, e.g., a sum of numerical single-camera promises. 
By tuning k, we control the relative weights of high/medium confidence votes. 
In the current implementation, we sets k=2. 
That is, \sys{} moves a cell to the green as long as \sys{} has collected two medium-confidence votes for it. 

The threshold parameters $P_{high}$ and $P_{low}$ hinge on the tradeoff between refining existing results and exploring for new results. 
A lower $P_{high}$ would eagerly put cells in the green category, postponing processing additional cameras for them until much later in the search process. 
By comparison, a higher $P_{high}$ would be more reluctant in turning cells green; 
\sys{} will only pause processing on them if evidence is strong. 

Given that \sys{} is a recall-oriented system to minimize human effort, 
we set $P_{high}$ high so that \sys{} continues spending resource on promising cells to refine their ranking.
This is because we expect users to only inspect the top few cells; 
it is thus vital to include true cells in this small range. 
Based on the same rationale, we tune $P_{low}$ to a low value to admit more cells to the gray category. 
% In our implementation, we set $C_{high}$ to be -.0.73, where 99\% of the vehicles with instances below d are true vehicles, averaged across the samples 
% We profile video clips offline to determine the thresholds. 
We set $P_{high} = 1/d_{short}$, where 99\% of the bounding boxes with feature distance shorter than $d_{short}$ belong to the same vehicle. 
% per Tiantu, below is difficult to explain.
% and $p_{high} = 1/d_{long}$, where 99\% of the bounding boxes with feature distance ... \Note{fix, complete this}
% $C_{low}$ to be -0.91, where feature distances of 99\% of true vehicles are shorter than 0.91. \Note{fix this}
% We will evaluate sensitivity to the thresholds in Section~\ref{sec:eval}.
We will evaluate their sensitivity.
\subsection{The search process}
\label{sec:query:1}

\paragraph{Stage 1: Initial sampling of cells}
\sys{} starts a query by sampling from all cells and processing one camera for each. 

Based on videos from the starter cameras, 
\sys{} recognizes distinct objects in each cell; 
for each recognized object, \sys{} derives their cluster centroids. 
% Without cell promise estimated at this time, \sys{} just processes cells in random order. 
\sys{} may prioritize cells if heuristics is available on which cells are more likely to contain the target object, e.g., from rush hours or busier traffic intersections.
As \sys{} uses a low video frame rate (1 FPS) tolerable to the clustering algorithm (Section~\ref{sec:clustering}), it cover starter cameras from all cells with the lowest total cost. 
As initial sampling is done without information of the input image, 
it can be executed at the ingestion, as will be discussed below. 

After processing the \prim{} cameras for all cells in the query scope, 
\sys{} has the initial categories of cells with cells in each category ranked by their promises. 

\paragraph{Choosing \prim{} cameras}
The choices matter as they set the initial direction for search. 
Ideally, they should be the cameras most likely to have captured the target from a viewpoint similar to the input image. 
In practice, one could exploit knowledge on camera deployment to help pick \prim{} cameras, e.g., by choosing the camera that has the most similar viewpoint with the input image. 
Without assuming such a priori, our base design follows simple heuristics: 
% picking cameras that have captured the highest number of distinct objects. 
picking cameras that has the highest density of distinct objects. 
The hope is that their chances of having captured the target are higher; 
if the target is captured, even from a different viewpoint than the input, the resultant bounding boxes would show a decent similarity to the input and thus a high promise to \sys{}. 
\sys{} profiles each camera's density of objects offline and picks the starter cameras ahead of query. 
We evaluate sensitivity of \primary{} camera choices in Section~\ref{eval:sensitivity}.

\begin{comment}
% -- xzl: ``blind'' means without prior knowledge...
The design of \primary{} cameras is cost efficient yet blind.
First, since the target vehicle's orientation and location is not revealed until query time, the \primary{} camera's viewport that captures the target vehicle might not be the best among all cameras.
Second, although cameras in the same camera groups shows high redundancy that almost all vehicles are shared among the group, there are rare cases that the the target vehicle is missed in the \primary{} camera while is presented in other cameras.
In this case, \stp s that contains the target vehicles might be voted as ~\textit{weak reject} before query execution, and is re-visited in comparatively lower priorities.
\end{comment}

% \paragraph{Elastic to ingestion processing}
\paragraph{Initial sampling at ingestion time}
% As \sys{} chooses \prim{} cameras independent of any input image, it can do so at camera deployment time. 
Independent of input, the initial stage can be executed before queries. 
Pre-processing at ingestion is optional and elastic.
The number of \prim{} cameras \sys{} can process depends on resources, e.g., the number of GPUs owned by \sys{}. 
\sys{} processes unprocessed \prim{} cameras when a query starts, and caches the results for subsequent queries on the same scope of videos (\S\ref{sec:opt}). 

We also consider a situation of ample resources available to ingestion. 
Is it worth pre-processing multiple \prim{} cameras per geo-group? 
Our experiments, as will be shown in Section~\ref{sec:eval}, suggest diminishing returns. 
% for instance, with 2$\times$ compute cost paid at video ingestion, 
% the delays in executing a query are only further reduced by less than 10\% compared to 80\% reduction performed by only one \prim{} camera from each geo-group. 
This is because a small number of \prim{} cameras properly chosen can yield sufficiently accurate cell promises and the initial ranking. 

%\subsection{Step 2: Incremental (Guided?) sampling}
%\subsection{Step 2: Adding cameras to cells}
\label{sec:query:2}

\paragraph{Stage 2: Incremental search}
Based on initial sampling of starter cameras, 
\sys{} may be undecided on putting on a cell in the green or red category: 
the \prim{} may completely miss the target vehicle; 
its viewpoint on the target vehicle may differ significantly from the input image; 
or the \prim{} may have just captured a different, but visually similar vehicle. 

Specifically, \sys{} picks the next cell as follows: 
if the highest ranked gray cell that still has unprocessed cameras, \sys{} processes one additional camera for it;
if such gray cells are already exhausted, 
\sys{} processes the highest ranked green cell that still has unprocessed cameras; 
if no such cells, \sys{} moves to red cells, in hope of finding target object instances in those cells missed out previously. 
After selecting the next cell and processing one additional camera for it, 
\sys{} updates the cell's category and re-rank the cells. 
The updated categories and ranks will be \sys{}'s basis for picking the next cell. 

%For the video footage from each camera in a specific \cube{}, the cameras votes for \textit{strong accept} and adds 2 to the votes if $d_{min} < T_{low}$; the cameras votes for \textit{weak accept} if $d_{min} > T_{high}$ and adds 1 to the votes; the cameras votes for \textit{weak reject} and adds 0 to the vote.

\begin{comment}
\paragraph{Examples}
\Note{seems unnecessary. may keep if S5 appears too short}
Figure~\ref{fig:query} shows an example:
\encircle{1} Beginning at the top of the list, if the \cube{} is voted as \textit{strong accept} by the \primary{} camera, \sys{} accepts these \cubes{} and skips processing on those \cubes{}.

\encircle{2} For \cubes{} tagged by the \primary{} camera as \textit{weak accept}, \sys{} further investigates the vehicle similarity by processing video footage from on another camera \Note{how to pick this camera?}
If the new camera voted \textit{strong accept} or \textit{weak accept} for this \cube{} \Note{what's the semantics?}, 
\sys{} accepts those \cube{} as their count of votes is > 2.

Otherwise, the vote remains the same and requires further checking.
\encircle{3} For all pending \cubes, \sys{} incrementally samples the next camera and collect the votes.

% \sys{} conducts multiple rounds of incrementally sampling on all \stp s that currently has a vote $\leq 0$ from the prior round.
% \sys{} stops the sampling on \cubes{} when the number of cameras not visited in this \cube{} cannot turn vote count to 2, i.e., even if all the rest of the cameras votes \textit{strong accept}, the final vote is still $\leq 2$.
\end{comment} % 2 pg
    % !TeX root = main.tex

% -- below useful, but not critical --- 
\begin{comment} 

\textbf{--- Implementation ----}
\label{sec:impl}
\paragraph{Object detection and feature extraction}
We use YOLOv3~\cite{yolo} as the object detector for vehicles and ResNet-152~\cite{resnet} as the re-identification classification model.
ResNet-152 is implemented in PyTorch 1.0 and trained on images from 329 different vehicles from 34,760 images from CityFlow~\cite{cityflow} and Cars~\cite{cars} dataset, with all vehicles used in Section~\ref{sec:eval} excluded.

\paragraph{K-prediction model training}
We trained our K-prediction regression model and implemented K-means clustering algorithm in scikit-learn~\cite{kmeans}.
\end{comment}

\section{Optimizations with extra knowledge}
\label{sec:opt}
We present the following add-on optimizations. 
They are based on assumptions that we intentionally left out from \sys{}'s base design. 
We will evaluate them in \sect{eval}. 

\paragraph{Picking starter cameras based on posture similarity}
If the quantitative postures of deployed cameras are known to \sys{}, e.g., as part of per camera metadata, \sys{} can pick starter cameras as ones having the most similar postures to the origin camera. 
The rationale is that if the target object is captured by starter cameras, a similar viewpoint will boost the camera's confidence. 
To do so, \sys{} needs to estimate the posture of origin camera, which is different in each query. 
One one hand, \sys{} may rely on human analyst to annotate the posture (one image per query); 
on the other hand, it may automate the estimation with vision operators proposed by active vision research. 

\paragraph{Sampling cameras with complementary postures}
In its base design, when \sys{} samples a secondary (or subsequent) camera for a cell, it picks a random one from the same geo-group. 
Such decisions can be more informed by camera postures. 
While still keeping the choices of starter cameras, 
\sys{} picks the next camera as the one that offers the most different viewpoint compared to the prior camera sampled for, i.e., the N-th camera is always the camera that has the largest viewpoint difference with the (N-1)-th camera.

\paragraph{Exploiting camera correlations across locations}
Our base design already exploits strong correlation among \textit{co-located} cameras. 
In case quantitative correlations \textit{across camera geo-groups} become available, 
e.g., learnt through profiling, 
% \sys{} can retrofit existing techniques~\cite{rexcam} to augment its strategy of ranking cells. 
\sys{} can augment its strategy of ranking cells accordingly. 
To do so, \sys{} requires spatial correlation, i.e., the portion of overlapped objects across  geo-groups, and temporal correlations, i.e., the time range in which an object is likely to reappear in other geo-groups. 
Once \sys{} identifies a high-promise cell A and moves it to the green category, 
\sys{} identifies the most correlated gray cells, defined as the cells expected to share the highest portion of overlapped distinct vehicles with A, as estimated from the geo-group correlations. 
\sys{} will search in these before all other gray cells. 
Note that \sys{} still adds cameras to them incrementally, as in the baseline design. 
%\Note{how does this work? need more details. like visiting nearby cells after we output a cell to the user?}

% keep this? 
% If the input image comes with the location and time information, the alternative falls back to ReXCam~\cite{rexcam} that directly start the query from the cell that captured the input image, which significantly simplifies the problem by skipping the search the first occurrence. However, for \sys{}, we do not make such strong assumptions.

\paragraph{Reusing states of previous queries}
\sys{} speeds up a query's execution by reusing the states from prior queries on the same videos. 
These queries could be fully or partially executed. 
Their states include all distinct objects and their features from the starter cameras, and some of bounding boxes, distinct objects, and features from the remaining cameras. 
To the end, \sys{} reuses the existing distinct objects as the ``free'' estimation of the initial cell ranking; in incremental search, \sys{} may favor cameras for which partial results already exist. We will evaluate the former idea experimentally.  
    % !TeX root = main.tex

\section{Evaluation}
\label{sec:eval}
We answer the following questions in evaluation:

\noindent 
\S{\ref{eval:e2e}}
Can \sys{} achieve good accuracy with low delays? 

\noindent  
\S{\ref{eval:design}}
Are the key designs useful? 

\noindent
\S{\ref{eval:sensitivity}}
How is \sys{} sensitive to its parameters and query inputs? 

\noindent 
\S{\ref{eval:tradeoffs}} 
How beneficial is processing at ingestion time? 

\noindent 
\S{\ref{eval:knowledge}} 
%How does \sys{} perform with prior knowledge, e.g., camera orientation or correlations?
How effective are \sys{}'s add-on optimizations?

\subsection{Methodology}
\label{eval:method}

%\Note{a table to summarize the dataset? num of cams per location}
% !TeX root = main.tex

\begin{table}[]
    % \scriptsize
    \footnotesize
    \centering
    \includegraphics[width=0.49\textwidth]{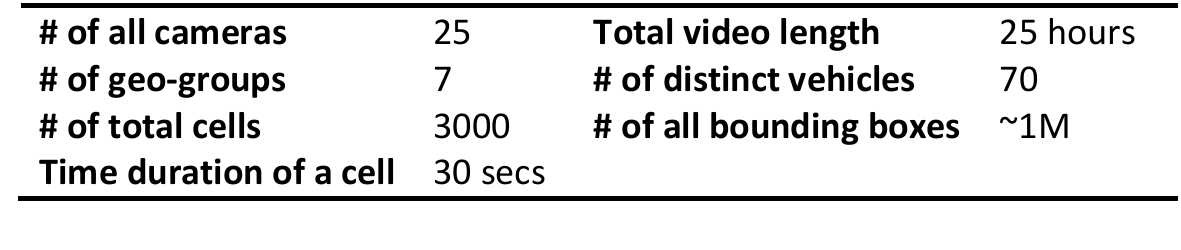}
    \caption{The augmented video dataset used in evaluation}
    \label{tab:dataset}
    \end{table}

\paragraph{Video Dataset}
%\Note{say this is the best dataset we can find. 
%Also mention it is difficult to find large multiple camera videos}
An ideal video dataset for benchmarking \sys{} would: 
(1) consist of long videos produced by many cameras; 
(2) come from real-world deployment and capture spatiotemporal patterns of vehicles; 
(3) have vehicle labels as the ground truth for accuracy evaluation. 
Real-world videos are preferred over simulators, e.g., VisualRoad~\cite{visualroad}:
while simulators can generate long traffic animations with multiple viewpoints, 
the resultant bounding boxes are ultimately based on vehicle motions and camera postures specified by us; 
it is unclear how well they reflect ReID in the real world, e.g., rarity of target objects, diverse cameras, and transient disturbance.

Unaware of such datasets in public, 
we use CityFlow~\cite{cityflow} published by NVIDIA for use in the AI City Challenge 2019. 
The dataset consists of 5 scenarios, from which we select the largest one (scenario 4). 
The scenario consists of 25 cameras at 7 traffic intersections (hence 7 geo-groups) of a northern American city. 
The scenario includes 30 minutes of videos, capturing 17,302 vehicle bounding boxes belonging to 70 distinct vehicles. 
We downsample videos to 1 FPS, a low frame rate adopted in prior video systems~\cite{focus, noscope, vstore}.

\begin{figure}
	\centering
	\includegraphics[width=0.48\textwidth]{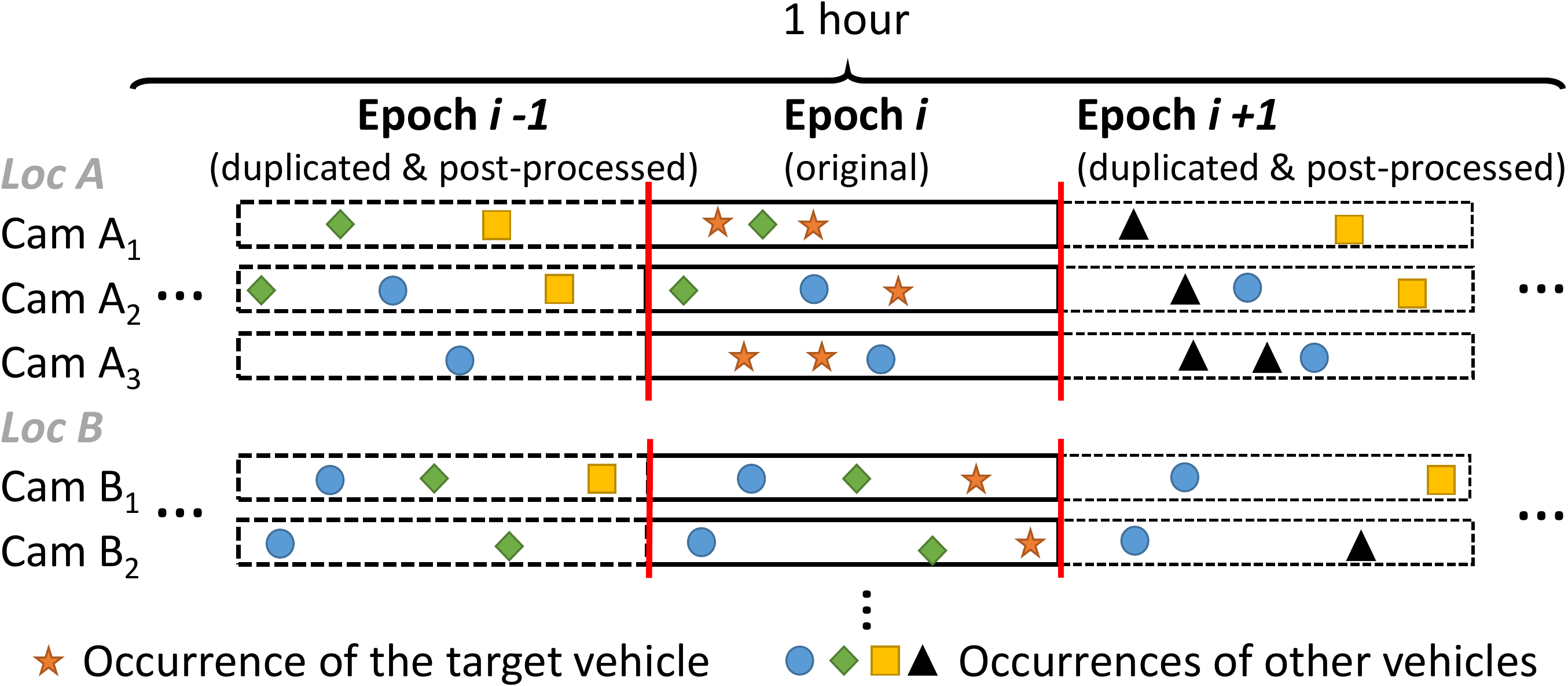}
	\caption{\textbf{Augmenting real-world city videos~\cite{cityflow} as our test dataset}: duplicating the original epoch; erasing random vehicles from each epoch; erasing the target vehicle from all but the original epoch.}
	\label{fig:dataset}
\end{figure}

We overcome a key shortcoming of the CityFlow dataset: 
all videos are short, each lasting around 30 seconds on a camera. 
% Short videos will largely simplify the query as the target object is not ``rare'' any more, as the number of \cubes{} are not sufficient. 
We therefore augment the dataset to extend video length. 
To do so, we make sure to:
(1) preserve the vehicle spatiotemporal patterns, within and across camera geo-groups; 
(2) keep the bounding boxes of target vehicles rare. 
Our augmenting procedure is shown in Figure~\ref{fig:dataset}. 
First, we extend each camera's video by duplicating the original video clip as many ``epochs''. 
Each epoch embraces videos clips from all cameras; 
within one epoch, the original spatiotemporal patterns are preserved. 
% Note that since the starting times of some original video clips are misaligned, 
% we pad them with small video clips of the same street scenes without any vehicles. 
Second, we remove a random fraction (0--1) of vehicles from each epoch, erasing their bounding boxes from all the video clips in that epoch. 
This diversifies the augmented videos over time, preventing them from becoming repeated loops of the original clips. 

% First, in each location, to best reserve the temporal correlations between vehicle appearances across cameras, we pad empty clips before/after each video clip to align all cameras to the same chronological scale.
% Second, to keep the rarity of the target vehicle for each query,  we only keep the target vehicle in the original video clip (dark blue) and remove them from synthetic clips (light blue).
% Third, we diversify the whole dataset by randomly dropping irrelevant vehicle instances in synthetic clips.

%Second, the videos are not well-organized as the start/end time of each video has time shifts of up to a few minutes. \Note{the 2nd is trivial}
%To tackle the above challenge, we manually duplicated each video footage and padded based on the original video footage.

We further ensure that target objects are difficult to find throughout all videos. 
For a query with an input image of target X, 
we exclude the origin camera from the query scope; 
we erase all X's bounding boxes from duplicated epochs while only keeping ones in the original epoch. 

The final videos used in evaluation are summarized in Table~\ref{tab:dataset}. 
It span 25 hours of videos, one hour per camera. 
% which are 12 epochs in total,
Together, the videos consist of 3000 cells, each lasting 30 seconds; 
the videos include more than 1 million bounding boxes. 
Given a query, 
only 243 (0.02\% of all) bounding boxes on average belong to the target vehicle, 
and 1.6 (0.5\% of all) cells on average contain the target. 

\paragraph{Query setup}
We test \sys{} on 70 queries, each for one distinct vehicle in the video dataset. 
A query contains one vehicle image randomly selected from all bounding boxes of the vehicle in the dataset. We then exclude the origin camera from the query scope. 
As described in \sect{overview}, a query carries no metadata, e.g., the timestamp or the origin camera of the input image, as opposed to prior work~\cite{rexcam,jain18arxiv}. 
%As the video footage ingests, \sys{} preprocesses videos from each \primary{} camera on time windows of 30s. 

%The pre-trained K-prediction model as shown in Equation~\ref{regression} has a parameter of $\mathbf{a}$ = [0.0298, 0.1047, 0.8737, -4.6762, -4.6512] and $b$ = 14.6888. 

\paragraph{Environment} % with 64GB DRAM
\sys{} runs on a 12-core Xeon E5-2620 v3 workstation with a NVIDIA Titan V GPU.
\sys{} runs YOLO~\cite{yolo} to detect vehicles and ResNet-152~\cite{resnet} to extract features.
We train ResNet-152 on images from 329 different vehicles from 34,760 images from CityFlow~\cite{cityflow} and Cars~\cite{cars} dataset, with all vehicles used in the evaluation excluded.

\paragraph{Accuracy Metric}
We evaluate query accuracy with \textit{recall at k}, 
% i.e., the percentage of all true cells (i.e. containing the target object) included in the top k cells as ranked by \sys{}. 
i.e., the fraction of all true cells (i.e. containing the target object) that have been included in \sys{}'s top k output cells. 
\textit{Recall at k} is commonly used for measuring accuracy of recall-oriented retrieval focusing on rare positives~\cite{rank, choppy}. 
By setting k as low as 5, the resultant metric (\textit{recall at 5}) 
measures the usefulness of query results when used with low human efforts, i.e. when a user reviews the top 5 cells returned by \sys{}. 
A high value of \textit{recall at 5} means that \sys{} successfully returns most if not all true cells to the user, 
since the true cells of most queries (> 98\% in our dataset) are fewer than 5. 

% Recall@5 not only evaluates the recall, but also the quality of the ranks \Note{why?} 

\paragraph{Speed metric}
As \sys{} keeps refining the rank of cells,
we report the times since a query's start until the output accuracy reaches a set of accuracy goals: 0.25, 0.50, 0.75, and 0.99. 

% The accuracy metrics used in the evaluation are precision, i.e., how many \stp in the list are true, and recall, i.e., how many true \stp are returned in the output list.
% \begin{myitemize}
% \item Precision: the number of 
% \item Recall: how many true trips are found
%\item Precision: How many vehicles are TP in the cluster;
%\item Recall: No. of target vehicles in best cluster/No. of target vehicles in the camera
%\item Trips
%(1) Number of trips found
%(2) True trip rankings
% \end{myitemize}
% !TeX root = main.tex

\begin{figure*}[ht]
\centering
    \includegraphics[width=0.97\textwidth]{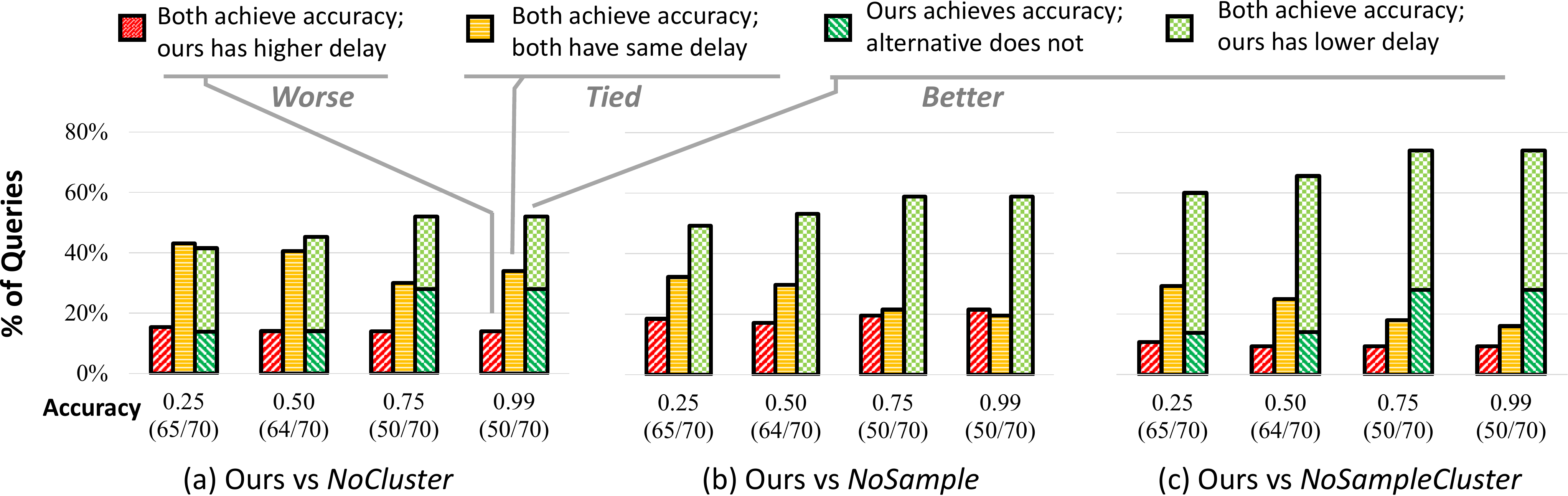}
    \caption{\textbf{Query-by-query comparison between \sys{} and the alternatives}, broken down by per-query comparison outcomes. Numbers on bottom: accuracy goals; (X/Y): X = number of queries that \sys{} reached the accuracy; Y = total query count}

\label{fig:e2e-quali}
\end{figure*}

%\begin{figure*}[ht]
%\centering
%    \includegraphics[width=0.99\textwidth]{figs/eval/e2e-exclude/e2e-quant-exclude.pdf}
%    \caption{Quantitative end-to-end performance comparisons}
%\label{fig:e2e-quant}
%\end{figure*}

\subsection{End-to-end performance}
\label{eval:e2e}
% \item Fig 9: we do not win much in (b). explain? Swap (c) and (b)? Where is `lost (accuracy)' in the figures? Do I miss them?

\paragraph{Accuracy}
\sys{} achieves high accuracy for most queries. 
% (i.e., putting the cells containing the target vehicle in the top-5 of the final rank. 
All 70 queries achieve an average accuracy of 0.87. 
Among them, 65 queries (93\%) meet or exceed an accuracy of 0.50; 
50 queries (70\%) meet or exceed an accuracy of 0.99. 
% Such accuracy is higher than what state-of-the-art reID algorithms achieved on individual bounding boxes. 
Such accuracy is higher than what can be achieved on individual bounding boxes, as will be shown in Section~\ref{sec:clustering}.
This validates our clustering approach: \sys{} can make robust decisions on cells based on fuzzy bounding boxes. 

We manually inspect the six queries where accuracy is low (< 0.5), 
attributing the root cause as the limitation of today's feature extractors. 
For instance, for a query with input vehicle 262, 
it is challenging even for humans to associate the input image with all the 30 true bounding boxes; 
not surprisingly, their features show long distances to the input. 
% It is also validated that \sys{} always shows higher accuracy than an alternative approach that does no clustering but decides based on individual bounding boxes, as discussed in Section~\ref{sec:clustering}.

\paragraph{Delays}
\sys{} achieves accuracy goals with moderate delays. 
% i.e., quickly refining the output cell list and converging to the final list. 
In querying 25 hours of videos on our single-GPU machine, 
\sys{} takes 59.5 seconds on average (stddev: 156.2, 90\% percentile: 457.5) to reach an accuracy of 0.50, 
and 108.5 seconds on average (stddev: 194.9, 90\% percentile: 488.0) to reach 0.99. 
Roughly, this speed is 830$\times$ of video realtime, i.e., 4.3 seconds to perform ReID on each hour of videos. 
% this speed is for queries that reach .99. 
% \notettx{0.75 is always the same with 0.99, can skip}

% \sys{} only takes an average of 106.8s to converge to the highest accuracy for each query over 25 hours of videos.

\subsection{Validation of key designs}
\label{eval:design}
%We next validate \sys{}'s key designs by comparing to the baselines. 

\paragraph{Alternatives}
We compare \sys{} to the following alternatives.
\begin{myitemize}

\item \textit{NoCluster}: Clustering is turned off. 
The alternative ranks a cell based on the minimum pairwise distance between the input image and bounding boxes in the cell. 
With the rank, it searches in cells by sampling from cameras as \sys{} does. 
% same threshold parameters as sys

\item \textit{NoSample}: Camera sampling is turned off.
The alternative randomly picks starter cameras for each camera group. 
It clusters bounding boxes and ranks cells accordingly, just as \sys{} does. 
Unlike \sys{} which adds one camera to a cell and updates the cell rank,
the alternative processes all cameras (in random order) for a cell before updating the rank. 

\item \textit{NoSampleCluster}: Both clustering and sampling are off. 
The alterative ranks a cell based on the minimum distance between the input image and bounding boxes; 
it processes all cameras for a cell before updating the cell rank. 
% \item \textit{RexCam}: In the spirit of RexCam~\cite{rexcam} that gradually returns the bounding boxes containing the target vehicles by exploring the spatio-temporal neighbors of the input image.
% However, ReXCam assumed known sources of input vehicle images, e.g., the input image comes from camera $cam$ at timestamp $t$.
% We eschew this assumption and by adding an exhaustive searches at the beginning of the query: randomly picking a bounding box image from all cameras and time windows until finding one promising bounding box image, i.e., $D (i, c) < T$. 
\end{myitemize}

%For baselines without adopting \sys{}'s incremental sampling strategy, i.e., \textit{naive} and \textit{\sys{}-NS}, they start with random choices of \primary{} cameras and visit all the rest of the cameras in each \cube{} rather than incrementally.
%
%For baselines without adopting clustering, i.e., \textit{naive} and \textit{\sys{}-NC}, we adopts the metric of minimum object feature distance between $f_{target}$ across all object feature rather than centroid feature distances.

%As show in Figure~\ref{fig:e2e-quali} and Figure~\ref{fig:e2e-quant}, we compare \sys{} with all other three baselines both qualitatively and quantitatively, respectively.
%\Note{these should go to methodology}

Figure~\ref{fig:e2e-quali} summarizes \sys{}'s competitiveness against the alternatives. 
On most queries, \sys{} outperforms the alternatives, either reaching higher accuracy or the same accuracy in lower delays. 
Only on a small fraction of queries \sys{} shows longer delays; \sys{} never fails to reach accuracy goals attainable to the alternatives. 
% We next present quantitative results. 

\paragraph{Clustering improves query accuracy}
\begin{comment}
By comparing against all three baselines, the outcome falls into three categories for \sys{}, i.e., won, tied, and lost, composed by five conditions, as shown below:
\begin{myitemize}
\item Lost (Accuracy): this refers to the outcome when \sys{} did not reach the target accuracy but the baseline did throughout the query.
\Note{by what time?}

\item Won (Accuracy): this refers to the outcome when the baseline did not reach the target accuracy but \sys{} did throughout the query.
\end{myitemize}
\end{comment}
%We carried out all 70 queries on \sys{} and three above competitive baselines and collects the traces of Recall@5 and the corresponding delay.
%As shown in Figure~\ref{fig:e2e-quali} (a)-(c), \sys{} significantly outperforms all baselines qualitatively across all accuracy goals, i.e., 0.25, 0.50, 0.75, 0.99, and \sys{} typically outperforms more on higher accuracy goals.
%Compared to the alternatives that match the input with individual bounding boxes,
\sys{}'s \textit{eventual} accuracy, i.e., the accuracy after processing all videos, reaches 0.87 averaged on all queries, 
while the alternatives (\textit{NoCluster} and \textit{NoSampleCluster}) reach 0.74 on average. 
The per-query accuracy gain is 0.13 on average (stddev: 0.28). 
Among all queries, \sys{}'s eventual accuracy is higher on 14 out of all 70 queries; 
on the remaining queries \sys{}'s accuracy ties with the above alternatives (mostly with short delays, see below) and is never lower. 
Clustering is vital in two ways: 
(1)  it is robustness against outlier bounding boxes and strong disturbance, the key to achieve high accuracy goals such as 0.99; 
% with \textit{NoCluster}, for example, up to 37\% queries fail to reach such accuracy goals. 
(2) based on clustering, \sys{}'s initial cell ranking is more accurate, ensuring speedy search.

\begin{figure}
	\centering
	\includegraphics[width=0.48\textwidth]{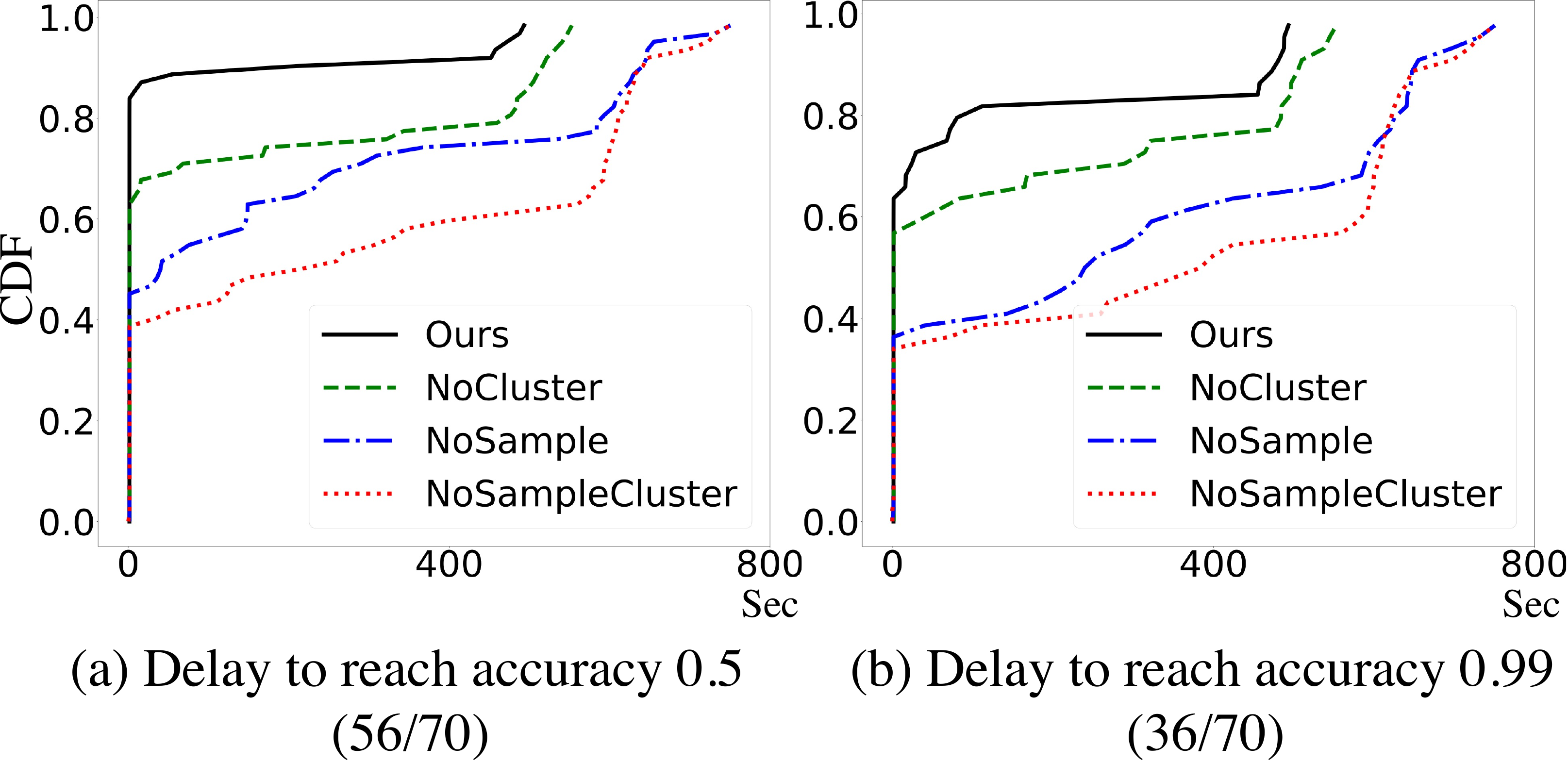}
	%\caption{Average delay to reach 0.50 and 0.99 across all baselines}
	%\caption{Average delay to reach 0.50 and 0.99 across all baselines}
	\caption{The CDF of query delays by \sys{} and the alternatives. (X/Y): X = the number of queries on which all the versions reach the accuracy goal; Y = total query count}
	\label{fig:delay}
\end{figure}

\paragraph{Camera sampling reduces query delays}
%As shown in Figure~\ref{fig:e2e-quant} (a)-(c), \sys{} significantly outperforms all baselines quantitatively across all accuracy goals, i.e., 0.25, 0.50, 0.75, 0.99.
%We evaluate the expected time \sys{} gain/lose, i.e., the product of the percentage of queries \sys{} won/lost and the delay differences.
We zoom in the queries and accuracy goals attainable to \sys{} and all the alternatives.
%The number of queries that either the baseline or \sys{} reached the accuracy milestone is tagged in Figure~\ref{fig:e2e-quali}. 
Figure~\ref{fig:delay} shows the delay CDFs.
With accuracy goals of 0.50 and 0.99, 
\sys{}'s delays are 2.9$\times$ and 1.7$\times$ shorter than \textit{NoCluster} on average, 
4.5$\times$ and 3.3$\times$ shorter than \textit{NoSample} on average, 
and 6.5$\times$ and 3.9$\times$ shorter than \textit{NoSampleCluster} on average. 
%\Note{percentage is more meaningful than absolute numbers}
%\Note{beside a weighted average, should also report things like histogram etc.}
%Compared with the \textit{Naive} (Figure~\ref{fig:e2e-quant}(a)), \sys{} is on average xx$\times$ faster across all accuracy goals, \sys{} is on average 223.9s-207.9s faster, while was only 18.6s-12.6s slower.
%Compared with the \textit{\sys{}-NC} (Figure ~\ref{fig:e2e-quant}(b)), \sys{} is on average xx$\times$ faster across all accuracy goals, \sys{} is on average 41.1s-69.8s faster, while was only 3.7s-8.6s slower.
%Compared with the \textit{\sys{}-NS} (Figure ~\ref{fig:e2e-quant}(c)), \sys{} is on average xx$\times$ faster across all accuracy goals, \sys{} is on average 134.9s-187.9s faster, while was only 27.4s-44.5s slower.
%\Note{percentage is more meaningful than absolute numbers}
The alternatives suffer from poor starter cameras which in turn result in poor initial cell ranking. 
% the poor ranking makes them process 6.5\% more videos \Note{quite minor} on average than \sys{} to reach the same accuracy. 

% !TeX root = main.tex

%First, as the choices of next camera \Note{what is this} is based on random sequences, the first (\primary{}) cameras might not be the most popular one and may missed up to more than 3$\times$ of target vehicles than \sys{} at ingestion time. 

\begin{comment}
\Note{The paragraph below might be useful, but not urgent. May remove}
\paragraph{Why did \sys{} occasionally underperform the baselines?}
Without clustering, baselines like \textit{Naive} and \textit{\sys{}-NC} rank \cubes{} based on the minimum individual feature distance that occasionally converges a true \cube{}'s rank earlier than \sys{} in less than 18\% queries.
The above observation is more popular on lower accuracy goals when ranking those easy \cubes{}.
\Note{the above sentence unclear. Rephrase}
With sampling, \sys{} occasionally accepts a \cube{} before it converges to the correct rank by further visiting additional cameras.
\sys{} will finally reach the same accuracy, though takes longer time until visited all the rest of the cameras in all \cubes{} after decisions are made on all undecided \cubes{}.
\end{comment}

\subsection{Sensitivity to parameters and inputs}
\label{eval:sensitivity}
\begin{figure}
\centering
\includegraphics[width=.99\linewidth]{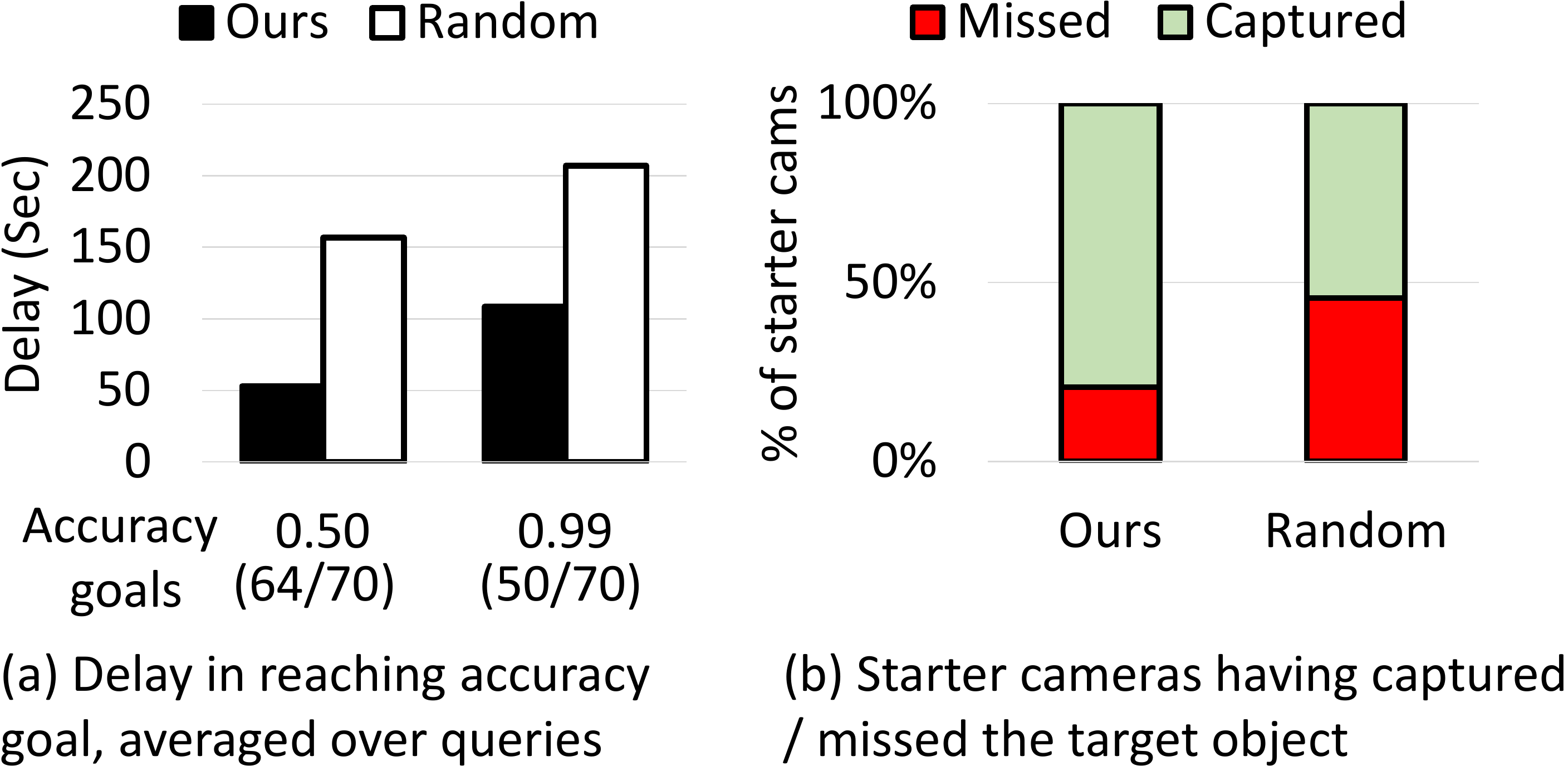}
%\caption{Breakdown of missed/unmissed objects in starter cameras and the overall delay}
\caption{\textbf{A comparison between \sys{}'s choices of starter cameras and random choices.} (X/Y) in (a): X = number of queries that \sys{} reached the accuracy; Y = total query count}
\label{fig:primary}
\end{figure}

\paragraph{Choices of \primary{} cameras matter}
While not affecting a query's eventual accuracy, 
the choice of starter cameras has a high impact on query delays. 
This is because the choice affects the initial rank of cells. 
% For instance,  if the \primary{} cameras missed a vehicle, the rank of this true \cube{} would be random, and thus \sys{} may waste hundreds of seconds on irrelevant \cubes{} until visiting the correct ones.  % USEFUL BUT VERBOSE
As shown in Figure~\ref{fig:primary}(a), with \primary{} cameras randomly picked,
the average query delays grow by 1.5$\times$ and 3.9$\times$ to meet accuracy goals of 0.50 and 0.99, respectively.
%the average delays grows by 1.5$\times$-3.9$\times$ by boosting from 38.4s and 109.2s to 149.8s and 169.8s for accuracy goals of 0.50 and 0.99, respectively.
Figure~\ref{fig:primary}(b) shows the cause: 
a substantial fraction of random starter cameras miss the target vehicle they should have captured (i.e. the target captured by other camera in the same geo-group), missing opportunity in optimizing the initial ranking of cells. 
% a substantial fraction of random starter cameras do not contain any bounding box of the target vehicle, resulting in poor ranking of cells. 
%\Note{check this}
% the target vehicle that are missed in \primary{} cameras grows more than 3$\times$ from 7.0\% to 22.5\%.
% !TeX root = main.tex

% % = (target vehicle missed but appeared in other cams in the group)/(groups that captured the target vehicle). those groups that do not catch the target vehicle are meaningless and wasn't included (edited) 

\paragraph{Moderate sensitivity to input images}
% \Note{is this stddev for the increase, or for the new value?} \notettx{std for the diff}
\sys{} shows resilience to different input images from our dataset. 
First, we replace the randomly selected input images with another random batch selected from the dataset. 
As a result, \sys{} sees an average of 0.04 difference in accuracy (stddev: 0.24); 
%\Note{for .5 and .99 accuracy levels?};
it sees delay differences of 7.7 seconds (11.7\%) and 13.4 seconds (12.5\%) for 0.50 and 0.99, respectively.
% and reduction in delays on average (std: 84.5 and 117.3). 
Second, we test ``easier'' input images by including the camera that produced the input image in a query's scope, while still excluding the input image from the scope.
As such, the query scope now has a camera with a viewpoint identical to the input image. 
\sys{} a sees moderate benefit: 0.05 higher accuracy on average (std: 0.15), 23.0\% and 12.1\% reduction in delays, and 1\% less processed videos on average.
% (std: 88.5s and 79.1s \Note{what are these}).
We attribute the results to our dataset characteristics: 
(1) Camera redundancy: given any input image, there are likely cameras offering similar viewpoints. 
(2) Decent image quality. 
Were the input images in poorer quality, e.g., with large occlusion or low resolution, they may confuse the neural networks used in \sys{}, resulting in lower accuracy overall. 

%First, vehicles usually appear in more than one location, so that the impact is comparatively small. 
%Second, in some cases, the \cubes{} are already accepted before processing the camera got removed.
%Third, due to camera redundancy that a vehicle is typically captured in multiple cameras, clustering can still effectively approximates distinct vehicles even if there is slight discrepancy in viewports, lightning conditions, etc.
%\input{fig-input}

\paragraph{Low sensitivity to thresholds} 
We learn $P_{high}$ and $P_{low}$ via offline profiling using the original video clips.
As the thresholds determine when to pause sampling cameras, 
their values may affect query delays but not the eventual accuracy.
With methods described in (\S\ref{sec:query}), we determine the default values as $P_{high}$=$1/d_{short}$=$1/0.73$ and $P_{low}=1/d_{long}=1/0.91$. 
%As shown in Figure~\ref{fig:thresholds}, the choices of thresholds have impacts on the performance while not significant.
We test \sys{} by deviating from such default values, i.e., $d_{short} \pm 0.1$ and $d_{long} \pm 0.1$. 
Across all 70 queries, the query delays only vary by less than 10\% on average.
The new thresholds increase delays on more than 90\% of the queries (average increase: 2.4 seconds); 
and reduce delays on the remaining (average reduction: 7.2 seconds). 
Based on the minor variation, we conclude that the default parameters are adequate; 
the benefit from fine tuning thresholds for individual queries are marginal. 
\subsection{Delay reduction by processing at ingestion}
\label{eval:tradeoffs}

% !TeX root = main.tex

\begin{figure}
\centering
\includegraphics[width=.99\linewidth]{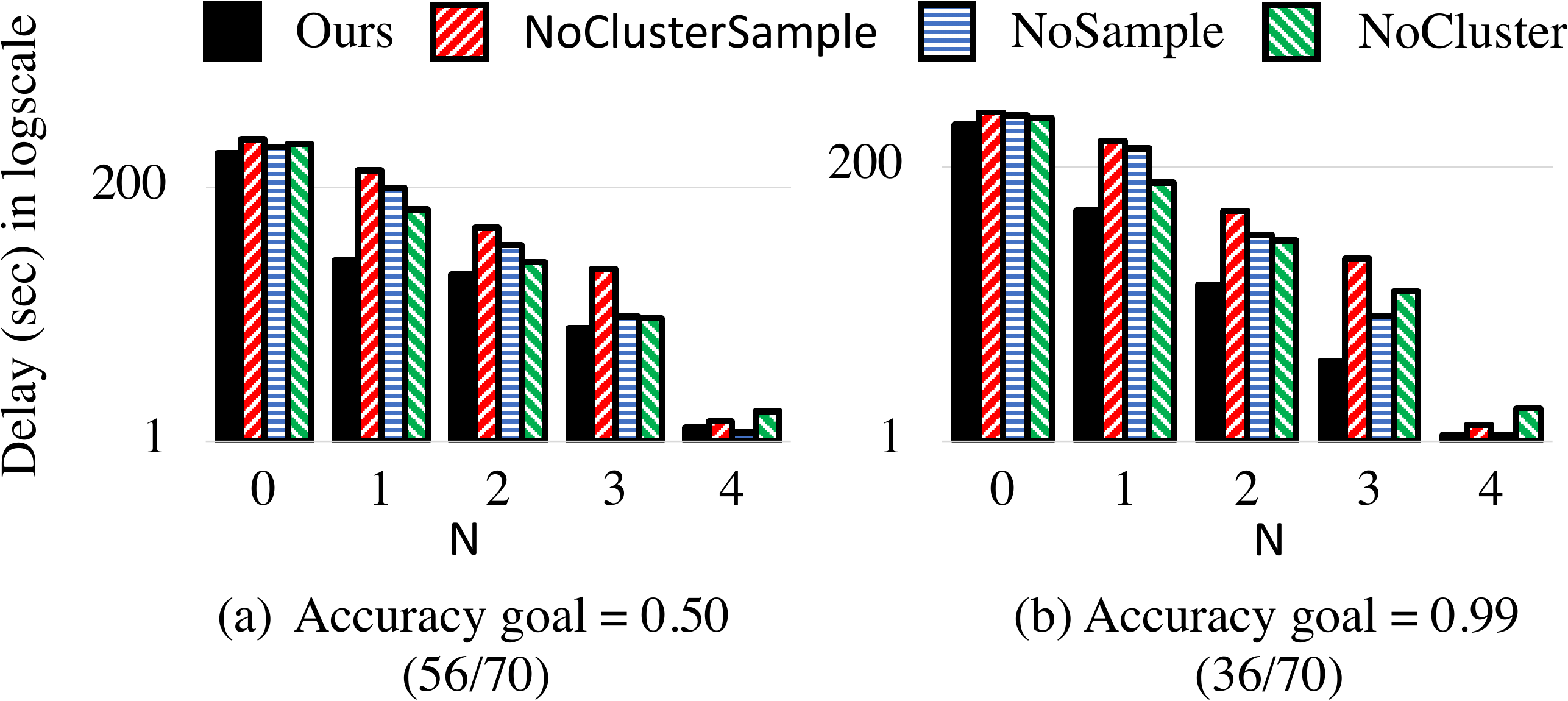}
\caption{\textbf{Query delays with $N$ cameras per geo-group pre-processed at ingestion time.} 
In case $N$ exceeds the total cameras of a geo-group, all the cameras are pre-processed. (X/Y): 
X = number of queries on which \textbf{all} versions reach the accuracy goal; Y = the total query count.
Y-axis in logscale.
%Problem: N from each group is different, some only has 2, some has 5}
}
\label{fig:tradeoff}
\end{figure}

% With more compute resource available at ingestion, \

Figure~\ref{fig:tradeoff} shows average query delays with a variety of \textit{N}, 
the number of starter cameras per geo-group pre-processed at ingestion time. 
With $N$ exceeding 4 (not shown in the Figure), \sys{} extracts all object features in real time, leaving only feature matching (negligible overhead) to query time. 

The results support lightweight preprocessing at ingestion.
(1) Pre-processing starter cameras reduces query delays substantially. 
Comparing to no pre-processing at all, pre-processing one starter camera per group reduces query delays by around 4$\times$. 
(2) Pre-processing more than 1 starter camera per geo-group yields diminishing returns, 
no more than 25\% delay reduction. 
(3) With the same pre-processing at ingestion time, 
\sys{} delivers much lower delays than the alternatives. 
% still beat \textit{NS-NC} by up to 6.5$\times$ and 4.2$\times$, beat \textit{NS} by up to 4.5$\times$ and 3.3$\times$, beat \textit{NC} by up to 2.9$\times$ and 3.8$\times$, on those queries that both \sys{} and the baseline reached 0.50 and 0.99, respectively.
% -- useful --
% \sys{}'s gain is not substantial when $N=0$, i.e., only process video data at query time, as the majority of time is spent on exploring the spatio-temporal coverage. 

 %86.3\% and 75.2\%.

% By processing more cameras at video ingestion, i.e., from 2-5, \sys{} only reduced 7.3\%-13.7\% and 16.0\%-24.8\% more delay than \sys{} on accuracy of 0.50 and 0.99, respectively.

%\begin{myitemize}
%\item \textit{No index}: An alternative design that only streams video to cloud disks at video ingestion and only processes the data during query time.
%\item \textit{Half index}: An alternative design that pays a little bit more cost by indexing a half of the cameras in each location a video ingestion.
%\end{myitemize}

\subsection{Impact of optimizations}
\label{eval:knowledge}
We evaluate the deployment-dependent optimizations (\S\ref{sec:eval})

\begin{figure}
\centering
\includegraphics[width=.99\linewidth]{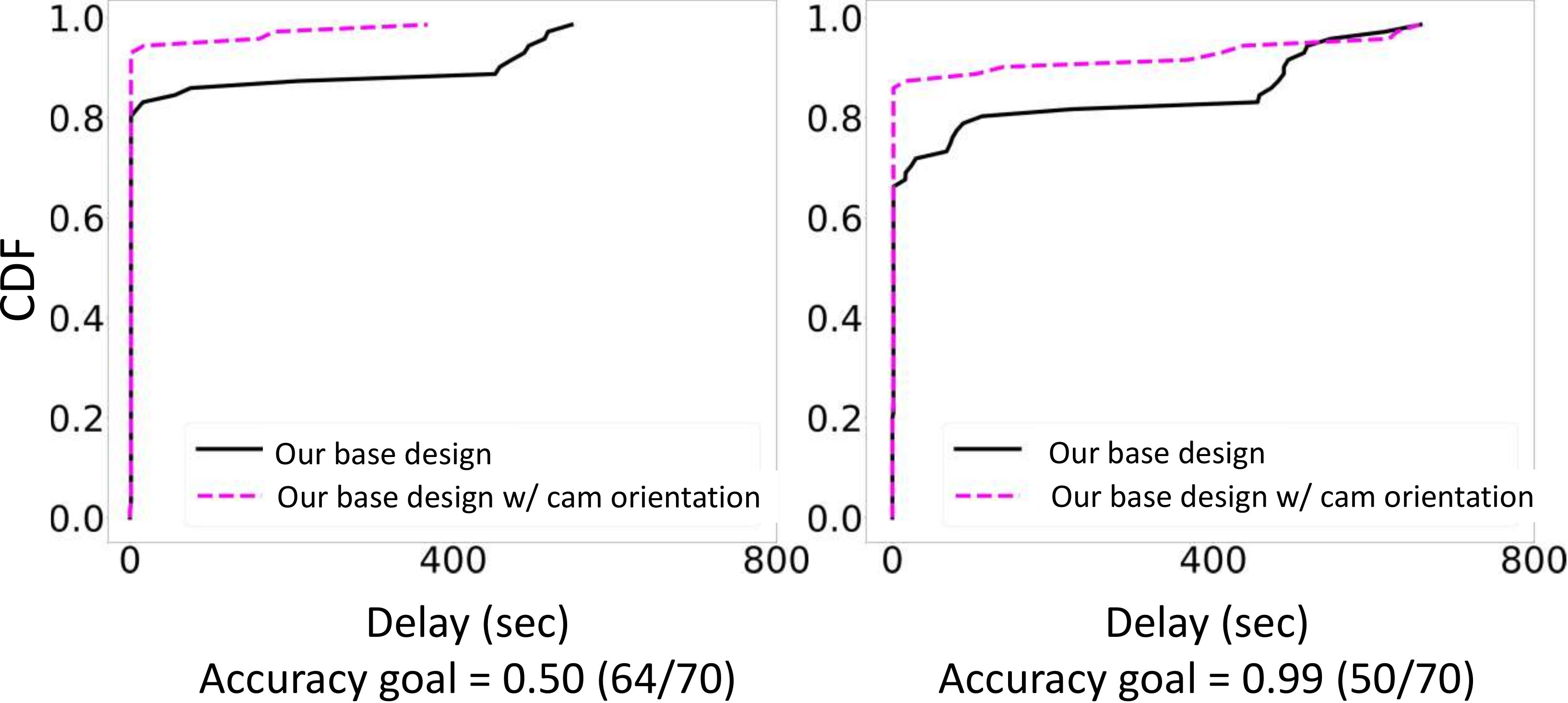}
\caption{\textbf{Delay CDFs of \sys{} augmented to exploit camera orientation knowledge}. (X/Y): X = number of queries that \sys{} reached the accuracy; Y = total query count}
\label{fig:starter-viewpoint}
\end{figure}

\paragraph{Picking starter cameras based on orientations}
We estimate deployed camera orientations from the map in Figure~\ref{fig:starter-viewpoint}. 
% While it is possible to extract viewpoints of input images with algorithms\cite{huang2019cvpr}, 
We use human-labeled viewpoints for input images of queries, minimizing inaccurate viewpoints. 
% We thus expect the results to be an upper bound of the benefit. 
Overall, this optimization tends to benefit queries used to be slow. 
On the queries used to have $\geq$70\% percentile delays
\sys{} sees on average 5.8$\times$ and 2$\times$ lower delays in reaching accuracy of 0.50 and 0.99, respectively. 
For the remaining 70\% queries, the delay reduction is negligible as most queries converge based on starter cameras.
%more moderate. (mean: XX, stddev: XXX)
The reason is that a significant better viewpoints from manually picked starter cameras have little impact on current well-performed queries, but on those queries that used to suffers from bad indexes. 
Besides, by replacing starter cameras that used to have a decent viewpoint, the initial rank of \cubes{} typically does not have sharp changes, so the query delay will not differ (those $<$70\% percentile). 
% On average, there is 82\% and 50\% reduction in delays to reach 0.50 and 0.99, respectively. 
% !TeX root = main.tex

\paragraph{Sampling cameras by complementary orientations}
With camera orientations of the dataset, 
\sys{} sees 4.6\% and 2.0\% reduction in delays to reach accuracy  of 0.50 and 0.99, respectively.
%\Note{we still beat te baselines by XXX...}
We identify two reasons for the minor benefit: 
co-located cameras are providing complementary viewpoints, and picking any of them is likely to help the search to similar degrees; 
the cameras with orientations opposed to the starter cameras can be inferior choices, e.g. capturing much fewer bounding boxes (and hence less likely the target object) than average cameras.

\paragraph{Exploiting camera correlations}
From the dataset, we manually derive the cross-location correlations. 
Assisted by the correlations, \sys{} reduces delays by 1.7\% and 1.9\% reduction to reach accuracy goals of 0.50 and 0.99, respectively.
The gain is insignificant due to the following reasons.
At the city scale, the correlations across locations are rather weak.
For instance, 
of any camera geo-group, only 2.4\%-14.3\% of the captured vehicles (mean: 8.4\%) are also captured by at least half of all other geo-groups. 
This is much weaker than camera correlation \textit{within} a geo-group:
in our dataset, 59\%-100\% of vehicles are captured by more than half of the cameras within the same group; 
it is also weaker than what was reported on smaller camera networks~\cite{rexcam}. 
As a result, the benefit from cross-location correlations is low. 
% By comparison, by sampling from starter cameras 
% In addition, our base design can yield a good enough ranking of grey cells; there is little room for improvement from fine tuning such a ranking. 

% for undecided cells that ranked high, \sys{} does not take significant longer time to process them, and the correlation metric only process them a little ahead of time.
% Second, a cell having high correlations with currently accepted cells never guarantee a good rank, as the rank output to the user is still based on the distance measure rather than human labels during offline profiling. 

\paragraph{Reusing states of previous queries}
%For the dataset, besides the starter cameras, in each cell, we further pre-processed a random number of cameras in 50\% randomly selected cells, before the query to emulate existence of warm data from prior queries.
% \sys{} can effectively speed up a new query by reusing intermediate state, notably recognized objects and their features, of previous queries.
\sys{} can effectively speed up a new query by reusing intermediate state of previous queries.
To show this, we test ten query pairs <$Q_{old}, Q_{new}$> on two accuracy goals of 0.50 and 0.99.
The input images are randomly picked, and are different within each pair. 
Within each pair: we run $Q_{old}$, terminates it once reaching the accuracy goal, and run $Q_{old}$ with the query state left from $Q_{old}$. Between pairs, we cleanse any query state. 
With $Q_{old}$ reaching accuracy of 0.50 (i.e. a ``brief'' query), 
the delays for $Q_{new}$ to reach accuracy of 0.50 and 0.99 are reduced by 86.2\% and 76.8\%, respectively. 
With $Q_{old}$ reaching accuracy of 0.99 (i.e., a ``thorough'' query), 
the delays for $Q_{new}$ are reduced by 86.2\% and 78.1\%, respectively.

\begin{comment}
First, we run query A and abort the query when it reaches an accuracy of 0.50 or 0.99.
We then run query B with another input image on existing processed data from query A.
In ten pairs of query A \& B we tested, from prior queries aborted at 0.50, \sys{} sees an average delay reduction by 86.2\% reach 0.50 on all queries \Note{how about delays in reaching .99?}; 
from prior queries aborted at 0.99, \sys{} sees an average delay reduction by 61.6\% on to reach 0.99 on 9 queries, with 1 query failed to reach 0.99 \Note{how about delays in reaching .50?}.
The significant savings come from existing distinct vehicle instances from those cells inherited from prior queries, and \sys{} no longer needs to detect objects or extract features on them.
\end{comment} % 4 pgs
%    \input{discussion}
    % !TeX root = main.tex

\section{Related Work}
\label{sec:related}

We discuss related work not covered previously. 

\paragraph{Optimizing video analytics}
% Extensive work has been proposed for various scenarios.
%To reduce the compute cost across multiple cameras, prior work~\cite{rexcam} profiled camera correlations and prioritize more correlated bounding boxes;
%ViTrack~\cite{vitrack} proposed  spatio-temporal compressive
%target detection;
%Caeser~\cite{caeser} encodes the camera topology and searches objects in distributed key-value stores.
%\sys{} generalizes the query execution without making assumptions on when and where did the input image come from, making the camera correlation model/topology in vain.
% The assumptions regarding camera redundancy, correlation, and query model for \sys{} targets larger cameras networks with tens of camera deployments and departs from the above work.
%To eliminate the redundancy and save DNN inference costs at query time, Focus~\cite{focus} clusters extracted feature vectors by cheap NNs during video ingestion.
%Differently, \sys{} uses clustering to improve inference accuracy.
% Above clustering heuristics are similar to hierarchical clustering rather than k-means clustering that considers the sum of squared distances from all data points. 
% We use k-means clustering because for re-id problems, the individual feature vectors could be more difficult to differentiate between visually similar objects.
Besides ReXCam~\cite{rexcam} and ViTrack~\cite{vitrack} discussed earlier, to reduce multi-camera inference cost, Caesar~\cite{caeser} encodes object activity correlation across cameras; 
Optasia~\cite{optasia} shares common work modules and parallelizes query plans; 
Jiang \textit{et al.}~\cite{jiang19hotedgevideo} initiate an abstraction of camera clusters to enable resource/data sharing among cameras.
There has been many works proposed to optimize video analytics with operator cascades~\cite{noscope,mcdnn,focus,shen17cvpr}, and format tuning to trade accuracy for cost-efficiency~\cite{vstore,videoedge,chameleon,videostorm,pakha18hotcloud}.
Focus~\cite{focus} saves cost by pre-processing videos with cheap NNs at ingestion.
Notably, it clusters object features to avoid redundant comparisons with target objects.
\sys{} uses clustering in a different way: to smooth out transient disturbances for higher ReID accuracy.
Extensive works are proposed to exploit collaborations between cloud and edge~\cite{lavea,deepdecision,filterforward,edgeeye,ravindran18hotedge};
cloud/edge and mobile devices~\cite{glimpse,cashier};
cloud and cameras~\cite{diva};
edge and cameras~\cite{vigil,reducto}; 
and edge and drones~\cite{wang18sec}.
Elf~\cite{elf} imposes energy planning for counting queries on resource-frugal cameras.
None above was designed for ReID queries over city-scale cameras.

\paragraph{Information Retrieval}
Recall-oriented retrievals, e.g., legal or patent search, is a group of tasks to find all relevant documents and a bad ranking typically incurs significant search efforts from domain experts~\cite{rank,choppy,pres}. 
As objects are rare in ReID tasks and typically requires domain knowledge in crime investigation/smart traffic planning, we position \sys{} to solve recall-oriented tasks, i.e., requiring all true cell to be retrieved, and adopt the metric of recall for evaluation.
%We evaluates \sys{}'s end-to-end performance by recall@5 that not only evaluates the recall, i.e., the ability to find all true \cubes{}, but also the search efforts from domain experts as only the top 5 \cubes{} are considered.

\paragraph{Reconstruction of object views}
To estimate the object view from another viewport, prior work~\cite{zhou17cvpr} makes attempts to estimate the per-pixel depth from a single image, but they are still far from recovering the full 3D volumetric representations from single images and cannot not explicitly estimate scene dynamics and occlusions.

    % !TeX root = main.tex

\section{Conclusion}
We built \sys{}, a practical object ReID engine that answers spatiotemporal queries.
\sys{} is built upon two unconventional techniques. 
First, \sys{} approximates distinct objects by clustering fuzzy object features emitted by ReID algorithms before matching with the input image. 
Second, to search in colossal video data, \sys{} samples cameras to maximize the spatiotemporal coverage and incrementally adds additional cameras on demand.
On 25 hours of city videos spanning 25 cameras, \sys{} on average reached an accuracy of 0.87 and runs at 830$\times$ video real time in achieving high accuracy.  % .75 pgs
    
    \bibliographystyle{plain}
    \bibliography{main}
%    \input{main.bbl}
    % That's all folks!
\end{document}